\documentclass[
 aps,
 prb,
 reprint,
 amsmath,
 amssymb,
 superscriptaddress,
 longbibliography,
]{revtex4-2}

\usepackage{graphicx}
\usepackage{times}
\usepackage{xcolor}
\usepackage{hyperref}
\usepackage{makecell}

\newcommand{\ket}[1]{\left| #1 \right\rangle}
\newcommand{\bra}[1]{\left\langle #1 \right|}
\newcommand{\braket}[2]{\left\langle #1 \middle| #2 \right\rangle}

\usepackage[ruled,linesnumbered]{algorithm2e}

\begin{document}

\title{Tensor Network Loop Cluster Expansions for Quantum Many-Body Problems}

\author{Johnnie Gray}
\email{johnniemcgray@gmail.com}
\affiliation{Division of Chemistry and Chemical Engineering,  California Institute of Technology, Pasadena, CA 91125, USA}

\author{Gunhee Park}
\affiliation{Division of Engineering and Applied Science, California Institute of Technology, Pasadena, CA 91125, USA}

\author{Glen Evenbly}
\affiliation{AWS Center for Quantum Computing, Pasadena, CA 91125, USA}

\author{Nicola Pancotti}
\affiliation{NVIDIA Corporation, 2788 San Tomas Expressway, Santa Clara, 95051, CA, USA}

\author{Eirik F.~Kj{\o}nstad}
\affiliation{Division of Chemistry and Chemical Engineering, California Institute of Technology, Pasadena, CA 91125, USA}

\author{Garnet Kin-Lic Chan}
\affiliation{Division of Chemistry and Chemical Engineering,  California Institute of Technology, Pasadena, CA 91125, USA}

\date{\today}

\begin{abstract}
We analyze the tensor network loop cluster expansion, introduced in [G. Park, J. Gray, and G. K.-L. Chan, Phys. Rev. B 112, 174310 (2025)] as a systematic correction to belief propagation, 
in the context of general quantum many-body problems. We provide numerical examples of the accuracy and practical applicability of the approach for the computation of ground-state observables for high bond dimension tensor networks, in two- and three-dimensions, with open and periodic boundary conditions, and for spin and fermion problems. We find that the contraction error converges approximately exponentially with cluster size, enabling accurate local observable and energy estimates for many systems where standard contraction methods are otherwise impractical.
\end{abstract}

\maketitle

%%%%%%%%%%%%%%%%%%%%%%%%%%%%%%%%%%%%%%%%%%%%%%%%%%%%%%%%%%%%%%%%

\section{Introduction}

Tensor networks (TN) provide a powerful variational ansatz for the study of strongly correlated quantum many-body problems. In one dimension (1D), the matrix product state (MPS)~\cite{Vidal2003, SCHOLLWOCK201196, CiracPerezSchuch2021} provides an efficient representation of ground states obeying the area law of entanglement, forming the basis for efficient algorithms, such as the density matrix renormalization group (DMRG)~\cite{white1992density, white1993density}. This framework has been generalized to higher dimensions, most notably through the projected entangled pair states (PEPS) formalism~\cite{verstraete2004renormalization, CiracPerezSchuch2021}.

One of the challenges for PEPS is the cost to approximately contract the tensor network~\cite{ORUS2014117, ran2020tensor, gray2021hyper, gray2024hyperoptimized}, for example, when computing ground-state energies or local observable expectation values. Considerable progress has been made in two-dimensional (2D) systems with open boundary conditions (OBC) or in infinite lattices~\cite{Jordaorus2008}. Beyond that, in more complicated cases such as with periodic boundary conditions (PBC) or for three-dimensional (3D) systems, accurate computation of observables at large bond dimensions remains challenging~\cite{gray2024hyperoptimized}. In these situations, the cost to locally optimize or evolve quantum states, for instance, with the simple update (SU) method~\cite{jiangAccurateDeterminationTensor2008a}, 
is usually much cheaper than the cost to compute observables by contracting the tensor network.

To mitigate this contraction cost, several cluster approximations have been introduced based on the SU gauge~\cite{Lubasch_2014, Lubasch2014prb, Jahromi2019, jahromi2020thermal, Vlaar2021}. At the same time, the relationship between the SU gauge~\cite{Ran2013,RanSuperortho2012}, and the techniques of belief propagation (BP) originally from the field of statistical inference~\cite{pearl2022reverend}, have been clarified~\cite{AlkabetzArad2021,pancotti2023one,Tindall2023Scipost}. As BP is exact on tree graphs, these works have rationalized the accuracy of the SU/BP approximation in general lattices in terms of the magnitude of loop correlations.
Belief propagation has been increasingly employed as a way to approximately contract tensor networks and these methods have notably achieved success in simulating some recent quantum experiments~\cite{Begusic2024, Tindall2024ibm, Orus2024ibm}. 
But to go beyond the BP approximation, it is  necessary to account for the missing loop correlations. One way to do so was introduced in Ref.~\cite{evenbly2025loopseriesexpansionstensor}, where a rigorous loop series expansion was introduced to systematically incorporate corrections to the  BP estimate.

The current work is concerned with an alternative, and in some ways simpler, approach to improving on the BP approximation based on a loop cluster expansion (see Fig.~\ref{fig:schematic}), which is available in our open source package \texttt{quimb}~\cite{gray2018quimb}. We introduced the loop cluster expansion in Ref.~\cite{park2025simulatingquantumdynamicstwodimensional} and demonstrated it in the computation of expectation values in a 2+1~dimensional tensor network. The method was inspired by the 
numerical linked cluster expansion (NLCE)~\cite{rigolNumericalLinkedClusterApproach2006, tangShortIntroductionNumerical2013a} and the
techniques of generalized BP (GBP)~\cite{yedidiaGeneralizedBeliefPropagation2000, yedidiaUnderstandingBeliefPropagation2003,yedidiaConstructingFreeenergyApproximations2005}, which both provide cluster expansions of the free energy, but our technique simplifies GBP by avoiding the use of generalized messages. Here we provide a more detailed description and analysis of the loop cluster expansion, and assess its performance for expectation value approximation in a range of models, in both 2D and 3D, with periodic and with open boundary conditions, and for spins and for fermions.

%%%%%%%%%%%%%%%%%%%%%%%%%%%%%%%%%%%%%%%%%%%%%%%%%%%%%%%%%%%%%%%%

\begin{figure}[t!]
    \centering
    \includegraphics[width=1.0\linewidth]{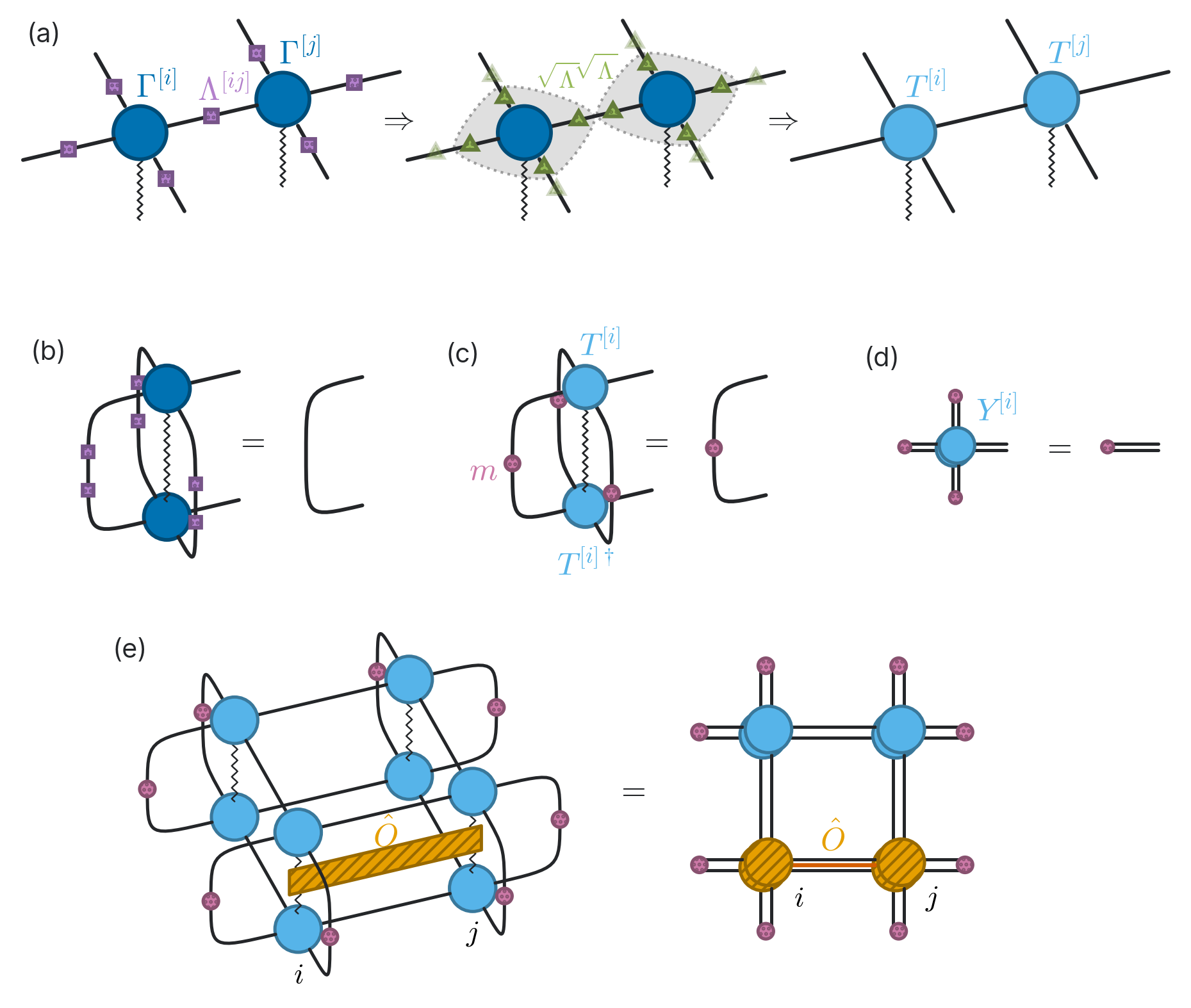}
    \caption{
    (a) Simple update (SU) gauging scheme with the Vidal gauge, and relation to the symmetric gauge.
    (b) SU local canonical condition. 
    (c) BP fixed-point condition.
    (d) Single layer or `top-down' view of the BP fixed-point condition.
    (d) The quantity $\langle \Psi | \hat{O}| \Psi \rangle_r $
    for a two-site observable $\hat{O}$ acting on sites $i,j$, evaluated with a single cluster $r$, and the equivalent single layer view.
    }
    \label{fig:subp_schematic}
\end{figure}

\section{Methods}

We start by briefly introducing the simple update (SU) and belief propagation (BP) for quantum tensor network states. Consider a tensor network representation of the (unnormalized) wavefunction $|\Psi\rangle$,
\begin{align}
     |\Psi\rangle = \mathcal{C} \left(\prod_i T^{[i]}\right),
     \label{eq:tn}
\end{align}
where $T^{[i]}$ are the tensors on the lattice, and $\mathcal{C}$  contracts common indices. 
In the SU gauging scheme, the wavefunction involves additional diagonal matrices $\Lambda^{[ij]}$ at the bonds between sites $i$ and $j$,
\begin{align}
     |\Psi\rangle = \mathcal{C} \left(\prod_i \Gamma^{[i]} \prod_{ij} \Lambda^{[ij]} \right),
\end{align}
which is also known as using the Vidal gauge \cite{Vidal2003, Vidal2004, Vidal2007} (Fig.~\ref{fig:subp_schematic}(a)). In this gauge, the tensors satisfy a local canonical condition~\cite{jiangAccurateDeterminationTensor2008a, RanSuperortho2012, Ran2013, Jahromi2019, Tindall2023Scipost}, shown in Fig.~\ref{fig:subp_schematic}(b). The form in Eq.~\ref{eq:tn} can be recovered by absorbing the square root $(\Lambda^{[ij]})^{1/2}$ into $\Gamma^{[i]}$ at each site $i$, i.e., $T^{[i]} = \Gamma^{[i]} \prod_j (\Lambda^{[ij]})^{1/2}$, called the symmetric gauge~\cite{Tindall2023Scipost, jiangAccurateDeterminationTensor2008a, AlkabetzArad2021}.

We can use BP to evaluate the tensor network norm $Z = \langle \Psi|\Psi\rangle$, where we use $Z$ to indicate the connection with  partition functions, the original setting for the application of BP.  $Z$ is a double-layer tensor network, which can also be viewed as a single-layer tensor network with tensors $Y^{[i]}~=~{T^{[i]\dagger}~\cdot T^{[i]}}$. The BP messages satisfy the BP fixed-point condition (Fig.~\ref{fig:subp_schematic}c). 
The single layer view is shown in Fig.~\ref{fig:subp_schematic}(d).
In the symmetric gauge above, the messages can be represented by
$m^{ij} = \Lambda^{[ij]}$. The SU canonical condition is then seen to be equivalent to the BP fixed point condition on the messages~\cite{AlkabetzArad2021, Tindall2023Scipost}. 

The BP approximation to $Z$ is obtained by contracting each tensor $Y^{[i]}$ with its surrounding messages (after normalizing the messages such that $m_{ij} \cdot m_{ji} = 1$) reducing each tensor to a scalar, and multiplying the scalars together, i.e.
\begin{gather}
    Z \approx \prod_i z_i \notag \\
    z_i = \mathcal{C}  \left(Y^{[i]} \prod_{j \sim i} m^{[ij]}\right)
\end{gather}
 Similarly, the free energy $F = \log Z = \sum_i \log z_i$. A local observable on site $i$,  $\langle \Psi |\hat{O}|\Psi\rangle / \langle \Psi | \Psi\rangle$ corresponds to the ratio of two tensor networks, which in this BP form clearly reduces to $o_i  / z_i$, where we introduce the notation $o_i = \langle \hat{O}^{[i]} \rangle_i$, and the latter indicates that the contraction is performed over the tensor at $i$ and its surrounding messages. Equivalently, we can define a generating tensor at site $i$, $Y(\lambda)^{[i]} = Y^{[i]} \cdot e^{\lambda O^{[i]}}$ (where $\cdot$ here indicates the values of the tensors are multiplied together) and derive the observable from the $\lambda$ dependent free energy, using $\langle \hat{O}\rangle = \frac{\partial}{\partial \lambda} \log Z(\lambda) \Big|_{\lambda=0}$, where $Z(\lambda)$ is the tensor network with the generator tensor. 

The BP approximation corresponds to an exact subset of the terms in the original TN sum.
This is because it arises from approximating the contraction over bonds in the (double-layer) TN by replacing an identity matrix on each bond $ij$, with $I \approx m_{ij} \otimes {m}_{ji}$.

\begin{figure}[t!]
    \centering
    \includegraphics[width=1.0\linewidth]{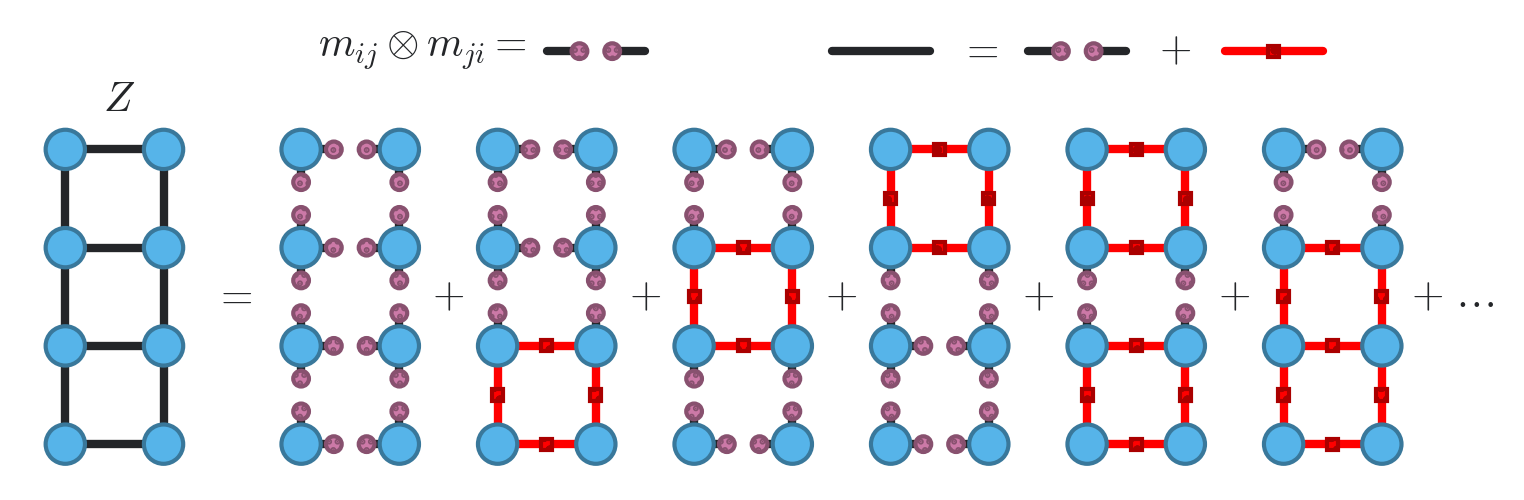}
    \caption{Partition function $Z$ computed as a loop series expansion. The BP approximation corresponds to approximating the contraction in $Z$ by inserting projectors (defined using the BP messages), denoted by the pair of purple circles, along the bonds, yielding the first term on the right hand side of the equation. Corrections to the BP approximation then consist of loops, where one evaluates the remaining contributions to the bond contractions in the space orthogonal to the projectors, denoted by the red lines. Non-loop contributions vanish due to the BP fixed point condition, shown in Fig.~\ref{fig:subp_schematic}(d).}
    \label{fig:loop_series}
\end{figure}

The exact $Z$ can be recovered by including the orthogonal contributions from $ I - m_{ij} \otimes {m}_{ji}$. Ref.~\cite{ChertkovChernyak2006, ChertkovChernyak2006b, evenbly2025loopseriesexpansionstensor} introduced an expansion around this mean-field in this ``excitation space''. Because of the BP fixed point condition, many contributions vanish, and the non-zero corrections can be depicted graphically in terms of loops on the TN graph, giving rise to an expansion in terms of loops. In this loop series expansion, the partition function can be obtained as a sum over all unique loop products on the lattice,
\begin{align}
    Z = \sum_\text{unique loop products} \prod_n z^\text{loop}_n
\end{align}
where $n$ enumerates all loops in a given loop product. 
A graphical example is shown in Fig.~\ref{fig:loop_series}.
Using the loop series to extract the free energy density, Ref.~\cite{evenbly2025loopseriesexpansionstensor} achieved a 3-4 orders of magnitude improvement over the naive BP expression for the free energy density in the thermodynamic limit.

In this work, instead of working with the loop series expansion, we consider cluster expansion approaches to improve the free energy. A simple way to improve the BP approximation is to perform an exact contraction over \emph{disjoint} clusters of sites. Dividing the lattice into such clusters, we then have
\begin{align}
Z = \prod_r Z_r
\end{align}
where $Z_r = \langle \Psi | \Psi\rangle_r$, and $r$ enumerates the disjoint clusters with their surrounding BP messages around them. Applying the same procedure to the definition of $\langle \hat{O}\rangle$, the expectation value is similarly obtained as $O_r = \langle \Psi | \hat{O} | \Psi\rangle_r / \langle \Psi | \Psi\rangle_r$. 
The numerator for a single cluster is shown in Fig.~\ref{fig:subp_schematic}(e).
This cluster approximation has been widely applied in the literature~\cite{Lubasch_2014, Lubasch2014prb, Jahromi2019, jahromi2020thermal, Vlaar2021, gaoFermionicTensorNetwork2025} to compute expectation values in TN with complex geometries where other approximate contraction methods are difficult to apply.

\begin{figure}[t!]
    \centering
    \includegraphics[width=\linewidth]{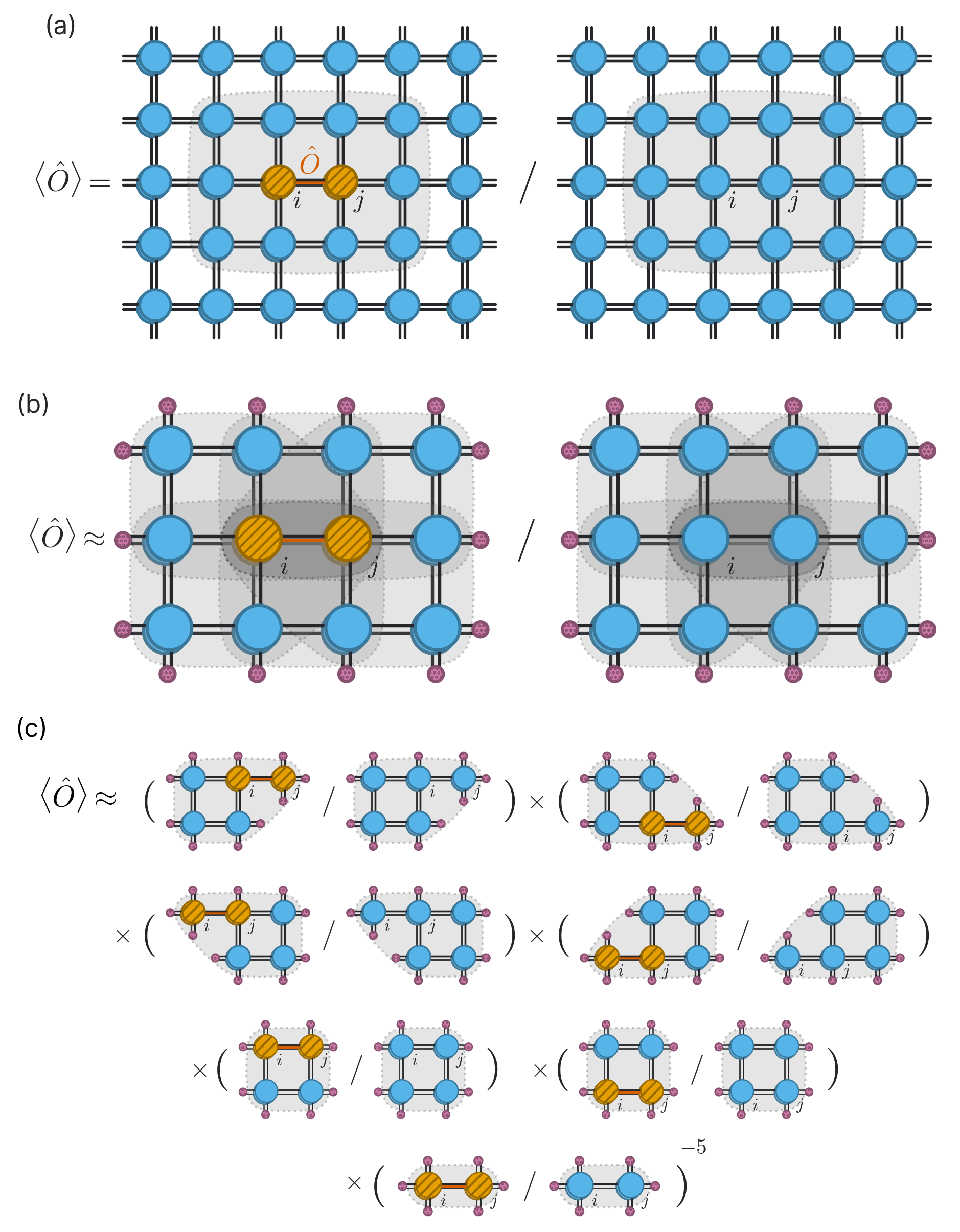}
    \caption{Overview of loop cluster expansion calculation of an observable $\hat{O}$ acting on two neighboring sites $i$ and $j$ (slashed orange sites connected by orange bond, see Fig.~\ref{fig:subp_schematic}(e)).
    (a) The exact expectation $\langle\Psi|\hat{O}|\Psi\rangle / \langle\Psi|\Psi\rangle$. Shaded region shows a local cluster, $r_{12}$, of $C{=}12$ sites.
    (b) Single cluster approximation of $\langle \hat{O} \rangle$ using the local cluster. Shaded regions show all clusters up to size $C{=}5$ that contain both target sites.
    (c) Loop cluster expansion (Eq.~\eqref{eq:product_cluster_expansion}) approximation of $\langle \hat{O} \rangle$ using all such local clusters.
    }
    \label{fig:schematic}
\end{figure}

\begin{algorithm}[t!]
\caption{Pseudo-code of loop cluster expansion computation of the energy corresponding to a sum of two-site contributions for a PEPS wavefunction $\Psi$, using the product-formula, Eq.~\eqref{eq:product_cluster_expansion}}
\label{alg:lce}
\KwIn{state $\ket{\Psi}$, Hamiltonian $\hat{H} = \sum_{\langle i,j \rangle} \hat{h}_{ij}$, max cluster size $C$}
\KwOut{Energy estimate $E$}
Converge BP messages $\{m^{ij}\}$ for $\braket{\Psi}{\Psi}$\;
$E \gets 0$\;
\ForEach{$\langle i,j \rangle$}{
    $\mathcal{R} \gets \{r : r \text{ is a loop cluster}, |r| \leq C, \{i,j\} \subseteq r\}$\;
    Close $\mathcal{R}$ under $\cap$: add $r_a \cap r_b\ \forall\, r_a, r_b \in \mathcal{R}$ until unchanged\;
    $c_r \gets 1 - \sum_{a \supsetneq r} c_a$, $\forall\, r \in \mathcal{R}$ in decreasing $|r|$\;
    \ForEach{$r \in \mathcal{R}$}{
        Let $r'$ be the largest loop $\subseteq r$ with equivalent contraction using BP fixed point condition (tree-like parts removed)\;
        \If{$r' \neq r$}{
            $c_{r'} \mathrel{+}= c_r$\;
            remove $r$ from $\mathcal{R}$\;
        }
    }
    $E_{ij} \gets 1$\;
    \ForEach{$r \in \mathcal{R}$ with $c_r \neq 0$}{
        $N_r \gets \braket{\Psi}{\Psi}_r$, $H_r \gets \bra{\Psi}\hat{h}_{ij}\ket{\Psi}_r$ with $\{m^{ij}\}$ on $\partial r$\;
        $E_{ij} \gets E_{ij} \times \bigl(\mathrm{contract}(H_r) / \mathrm{contract}(N_r)\bigr)^{c_r}$\;
    }
    $E \gets E + E_{ij}$\;
}
\Return{$E$}
\end{algorithm}

Motivated by the generalized BP (GBP)~\cite{yedidiaGeneralizedBeliefPropagation2000, yedidiaUnderstandingBeliefPropagation2003,yedidiaConstructingFreeenergyApproximations2005} approximation and the numerical linked cluster expansion~\cite{rigolNumericalLinkedClusterApproach2006,tangShortIntroductionNumerical2013a}, in Ref.~\cite{park2025simulatingquantumdynamicstwodimensional}, {we introduced a different way to systematically improve the BP result and applied it to the problem of computing quantum expectation values. This was based on combining the results from different clusters up to a given size $C$ with appropriate counting numbers. Because of the BP condition, the only contributing clusters are (generalized) loops, thus  
we refer to this as a \textit{loop cluster expansion.}}
Although the GBP approximation is a similar cluster expansion technique for the partition function and free energy, it introduces additional generalized messages beyond those used in BP itself, 
the size and complexity of which rapidly become prohibitive in the quantum setting.
The primary simplification of the loop cluster expansion compared to GBP is from the use of BP messages. We note that this same simplification has also appeared in the BP literature, under the name of the cluster-cumulant expansion \cite{welling2012}.

The computational procedure is as follows. All {loop} clusters are generated up to $C$ sites around some target sites.
Some of the clusters share overlapping regions,
and thus, from the viewpoint of the loop expansion, they include the same loop contribution multiple times. 
{To avoid double-counting the overlapping regions, we must also consider the clusters arising from all region intersections (which may not be loops) and assign a counting number $c(r)$ to each cluster with a region $r$ based on the inclusion-exclusion principle.}
This can be computed recursively as $1 - \sum_{a \supsetneq r} c(a)$ where the sum over $a$ runs over all other regions that $r$ is a proper subset of.
Note that while certain non-loop regions appear in the counting numbers, their contraction maps exactly to that of the largest loop region they contain, or the BP fixed-point if they are fully tree-like. 

The partition function can then be obtained as 
\begin{equation}
    Z \approx \prod_r Z_r^{c(r)},
    \label{eq:pfn_cluster_expansion}
\end{equation}
and the local observable expectation can be computed as a ratio of two partition functions, 
\begin{equation}
    \langle \hat{O} \rangle = \frac{\langle \Psi | \hat{O} | \Psi \rangle}{\langle \Psi | \Psi \rangle}  \approx \prod_r \left(\frac{\langle \Psi | \hat{O} | \Psi \rangle_r } {\langle \Psi | \Psi \rangle_r } \right)^{c(r)} = \prod_r O_r^{c(r)}.
    \label{eq:product_cluster_expansion}
\end{equation}
In the evaluation of the observable, there are two tensor networks (one for the numerator and one for the denominator), and thus two choices of BP messages. For example, to compute $\langle \Psi | \hat{O} | \Psi \rangle$, one could use the BP messages obtained from applying the BP algorithm to $\langle \Psi | \hat{O} | \Psi \rangle$, but another choice is to use the BP messages from the norm TN $\langle \Psi | \Psi \rangle$. The latter becomes convenient when computing multiple different observables, commonly needed when computing the energies of TN states, without the need to run the BP algorithm for each observable. However, the BP equation is not satisfied at the observable sites anymore. This means that certain non-loop contributions from `anomalous' clusters must be used.
Note that so long as the same messages are used for the numerator and denominator, all disconnected clusters cancel between them, reflecting the linked cluster property.

As an example, the computation of a single two-site observable $\hat{O}$ is illustrated in Fig.~\ref{fig:schematic} for a 2D lattice with $C=5$.
The exact expectation in (a) can be approximated by (b) corresponding to the `single cluster approximation', with messages inserted on the boundary $\partial r$ of region $r$.
In the loop cluster expansion, it is approximated by a product of overlapping cluster expectations as shown in (c). The counting number $c(r)=-5$ appears to avoid double counting contributions from overlapping regions.
In Algorithm~\ref{alg:lce}, we give the full pseudo-code algorithm for computing the energy of a given PEPS using the loop cluster expansion.

In the BP or single cluster approximation, computing the observable using derivatives of the free energy or as the ratio of partition functions is equivalent. This is not the case in the loop cluster expansion, because
the derivative gives
\begin{align}
    \langle \hat{O} \rangle &\approx \frac{\partial}{\partial \lambda} \log \prod_r Z_r^{c(r)}(\lambda) \bigg|_{\lambda = 0}  \nonumber \\
    &= \sum_r c(r) \times \frac{\partial}{\partial \lambda} F_r(\lambda) \bigg|_{\lambda = 0} \nonumber \\ 
    &= \sum_r c(r) \times \langle \hat{O} \rangle_r .
    \label{eq:sum_cluster_expansion}
\end{align}
In Ref.~\cite{park2025simulatingquantumdynamicstwodimensional}, we argued that Eq.~\ref{eq:product_cluster_expansion} for the observable corresponds to a weighted geometric mean and Eq.~\ref{eq:sum_cluster_expansion} is a weighted arithmetic mean, and observed that, in the model we were considering, they produced very similar results. 
Here, we refer to Eq.~\ref{eq:product_cluster_expansion} as the loop cluster product formula, and Eq.~\ref{eq:sum_cluster_expansion} as the loop cluster sum formula. We will compare their numerical performance below.

Finally, we briefly note two differences between the loop cluster expansion and the loop series expansion~\cite{evenbly2025loopseriesexpansionstensor}.
First, Eq.~\eqref{eq:pfn_cluster_expansion}, for any finite $C$, directly includes the contributions of all products of the disconnected clusters up to size $C$ in $Z$. In the loop series expansion, one would in principle need to include an exponential (in system size) number of loop corrections to achieve the same effect, and thus in Ref.~\cite{evenbly2025loopseriesexpansionstensor} we carried out an additional step to obtain these contributions implicitly when estimating the free energy.
However, when computing local observables using the messages from $\langle \Psi | \Psi \rangle$ as in Eq.~\eqref{eq:product_cluster_expansion} and Eq.~\eqref{eq:sum_cluster_expansion}, the disconnected contributions cancel between the numerator and denominator, which makes the loop series expansion and loop cluster expansion formalisms more comparable. 
A second practical difference particular to quantum expectations, is that in the loop series expansion there is an increased contraction cost associated with inserting the excitation projectors $ I - m_{ij} \otimes {m}_{ji}$ (Fig.~\ref{fig:loop_series}) into the double layer norm network. This can be expensive in larger clusters and especially in 3D, but is avoided in the current loop cluster expansion approach.

%%%%%%%%%%%%%%%%%%%%%%%%%%%%%%%%%%%%%%%%%%%%%%%%%%%%%%%%%%%%%%%%

\section{Results}

\begin{table}[t!]
    \centering
    \begin{tabular}{|c|c|c|c|c|}
        \hline
        $~~C~~$ &
        ~~ 2D scaling ~~ & 
        ~~ 2D $|\mathcal{R}_C|$ ~~ & 
        ~~ 3D scaling ~~ & 
        ~~ 3D $|\mathcal{R}_C|$ ~~ \\
        \hline
        $2$  &  $\mathcal{O}(D^5)$ &    1 &  $ \mathcal{O}(D^7)$ &    1 \\
        $4$  &  $\mathcal{O}(D^6)$ &    2 &  $ \mathcal{O}(D^8)$ &    4 \\
        $5$  &  $\mathcal{O}(D^6)$ &    4 &  $ \mathcal{O}(D^8)$ &   16 \\
        $6$  &  $\mathcal{O}(D^8)$ &   15 &  $ \mathcal{O}(D^9)$ &  106 \\
        $7$  &  $\mathcal{O}(D^8)$ &   38 & $ \mathcal{O}(D^{10})$ &  484 \\
        $8$  &  $\mathcal{O}(D^8)$ &  126 & $ \mathcal{O}(D^{12})$ & 2688 \\
        $9$  &  $\mathcal{O}(D^{10})$ &  346 & $\mathcal{O}(D^{12})$ & 13408 \\
        $10$ &  $\mathcal{O}(D^{11})$ & 1077 &  &  \\
         \hline
    \end{tabular}
    \caption{Computational scaling of the loop cluster expansion for a two-site observable and PEPS with bond dimension $D$ as a function of cluster size $C$. 2D and 3D refer to square and cubic lattices respectively. $|\mathcal{R}_C|$ denotes the number of regions of size $C$ that overlap the two nearest neighbor sites (only a few of which carry the most expensive scaling). We assume $D \gg p$ (the physical bond dimension).
    }
    \label{tab:scalings}
\end{table}

\begin{figure}[t!]
    \centering
    \includegraphics[width=1\linewidth]{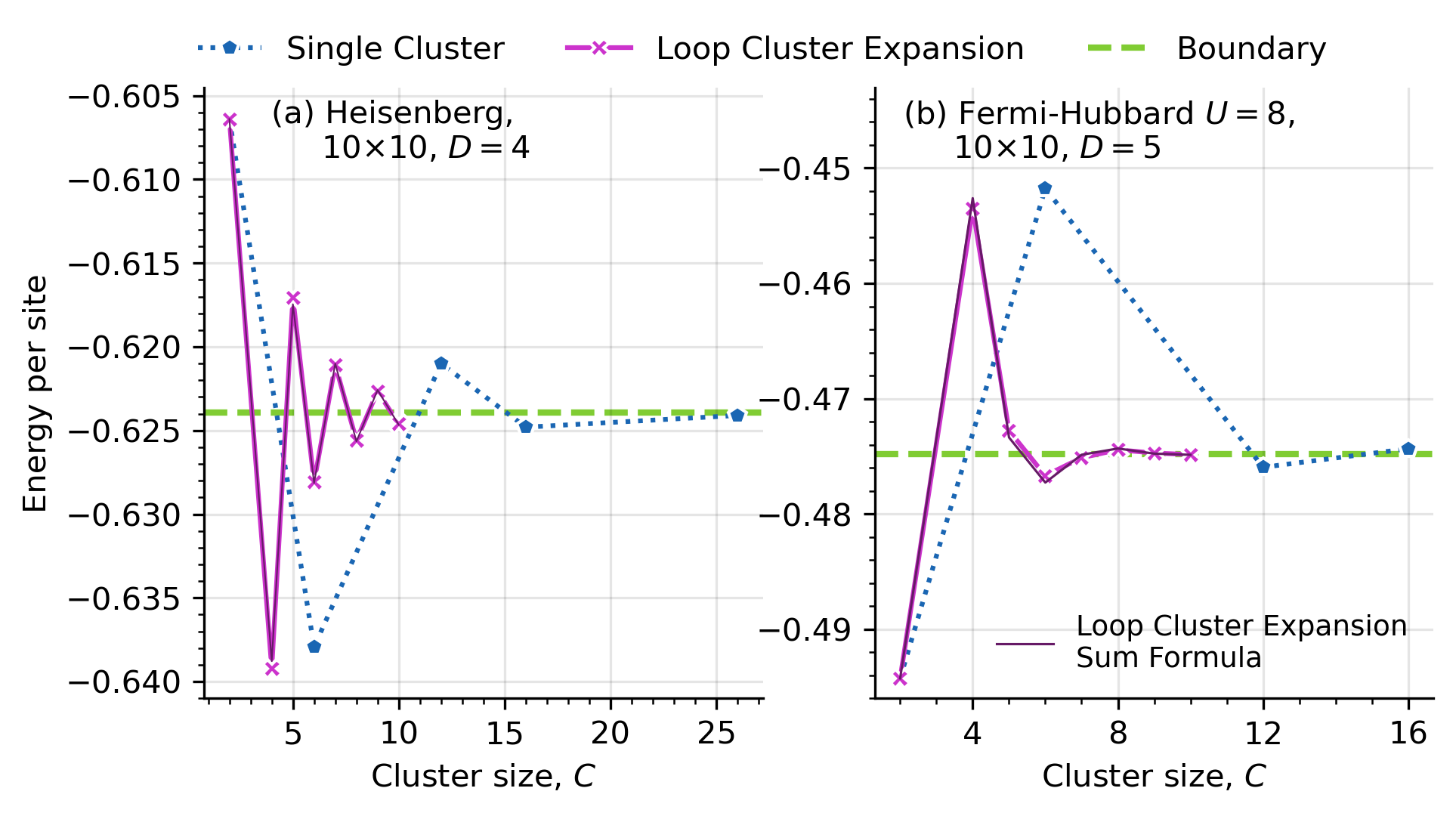}
    \caption{
    Example convergence of the loop cluster expansion for two square lattice OBC PEPS SU states on two different models, compared against the `single cluster' method and reference boundary contraction.
    The main loop cluster expansion data uses the product formula, whilst the thin darker line shows the (almost identical) sum formula for comparison.
    }
    \label{fig:energy-cluster-vs-expansion}
\end{figure}

We now illustrate the performance of the tensor network loop cluster expansions in quantum many-body ground-state problems. 
The accompanying code and data are available at~\cite{tn_loop_cluster_code_zenodo}.
We use the loop cluster expansions to calculate observables and report the error in the ground-state energy. We represent the ground-states as 2D or 3D PEPS and obtain approximate ground-states through the simple update (SU) scheme~\cite{jiangAccurateDeterminationTensor2008a}. For a given PEPS, two relevant errors exist: 1) the contraction error, an error in estimating the energy of the PEPS, and 2) the variational error, an error arising from the distance of the approximate PEPS to the exact ground-state. The purpose of the loop cluster expansion is to converge the contraction error.

We consider three models. 
Firstly the transverse field Ising model (TFIM):
\begin{equation} 
    H = J \sum_{\langle i,j \rangle} Z_i Z_j + B_X \sum_i X_i~,
\end{equation}
with Pauli operators $\{X_i, Z_i\}$ acting on site $i$, the sum $\langle i, j \rangle$ running over nearest neighbor sites, and $J{=}-1$ with field $B_X$ close to the critical point ($-3$ in 2D and $-5$ in 3D).
Secondly the Heisenberg model:
\begin{equation}
    H = J \sum_{\langle i,j \rangle} \left( 
    S^X_i S^X_j + 
    S^Y_i S^Y_j + 
    S^Z_i S^Z_j
    \right)
\end{equation}
with spin-1/2 operators 
$\{S^X_i, S^Y_i, S^Z_i\}$, and $J{=}1$ (anti-ferromagnetic coupling).
Finally the Fermi-Hubbard (FH) model at half-filling:
\begin{equation}
    H = -t \sum_{\langle i,j \rangle, \sigma} \left( c^\dagger_{i\sigma} c^{\vphantom{\dagger}}_{j\sigma} + c^\dagger_{j\sigma} c^{\vphantom{\dagger}}_{i\sigma} \right) + U \sum_i n_{i\uparrow} n_{i\downarrow}
\end{equation}
where $c^\dagger_{i\sigma}$ ($c^{\vphantom{\dagger}}_{i\sigma}$) creates (annihilates) a fermion of spin $\sigma \in \{\uparrow, \downarrow\}$ at site $i$, and $n_{i\sigma} {=} c^\dagger_{i\sigma} c^{\vphantom{\dagger}}_{i\sigma}$ is the number operator, with hopping strength $t{=}1$ and interaction $U=8$. 

We consider multiple geometries and boundary conditions, including the two-dimensional (2D) square lattice and the three-dimensional (3D) cubic lattice, with open or periodic boundary conditions (OBC or PBC).
In Table~\ref{tab:scalings} we show both the computational scaling of the loop cluster expansion as a function of cluster size $C$ and PEPS bond dimension $D$, as well as the number of new clusters, $|\mathcal{R}_C|$, that are introduced at each size $C$, for these two geometries.

\begin{figure}[t!]
    \centering
    \includegraphics[width=\linewidth]{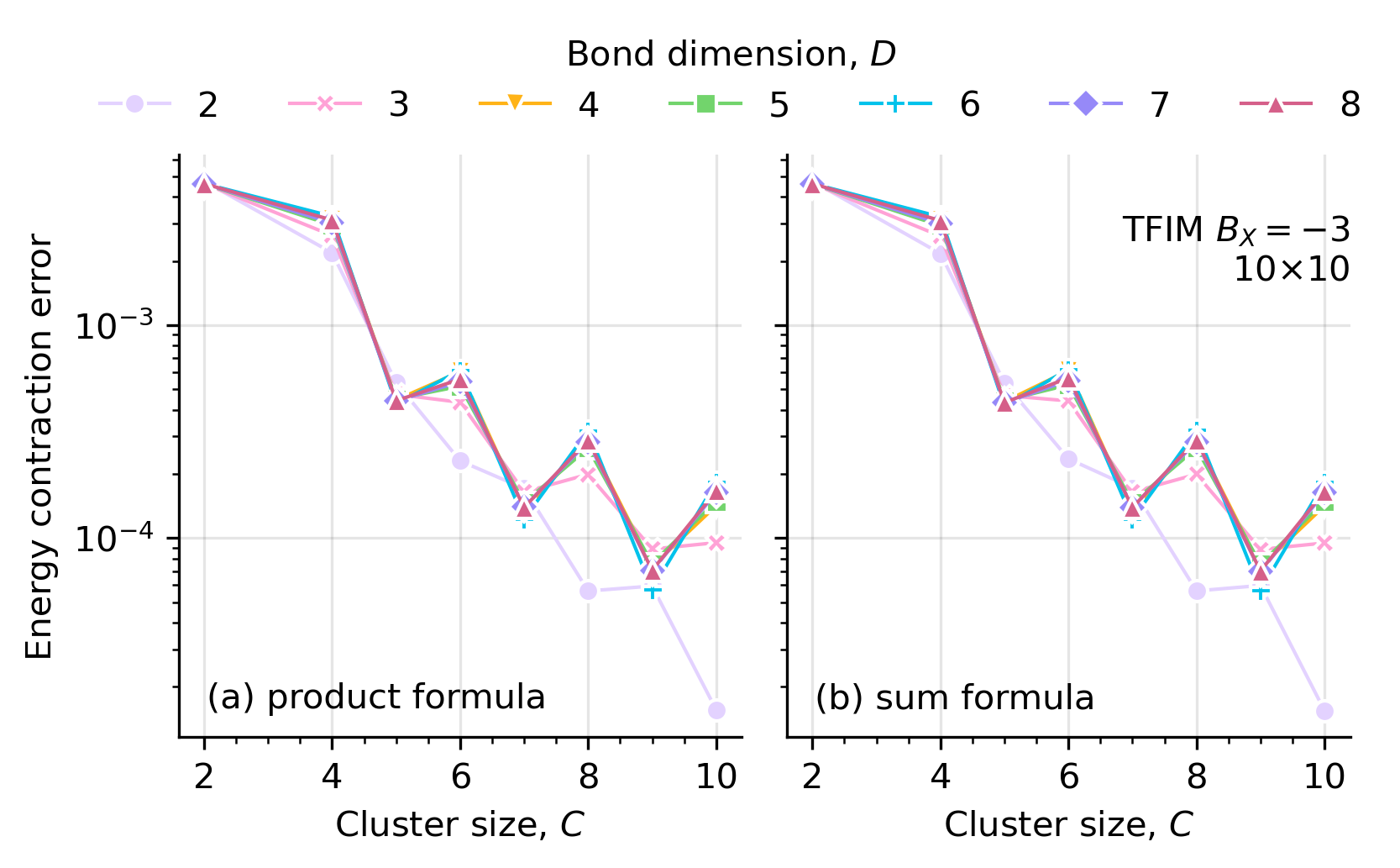}
    \caption{
    Relative energy contraction error for the 2D OBC TFIM with SU-optimized PEPS, as a function of cluster size $C$ and bond dimension $D$, computed with the loop cluster expansion (without extrapolation).
    (a) error using product formula, Eq.~\eqref{eq:product_cluster_expansion}.
    (b) error using sum formula, Eq.~\eqref{eq:sum_cluster_expansion}.
    }
    \label{fig:energy-contraction-error-square-obc-tfim}
\end{figure}

\begin{figure}[t!]
    \centering
    \includegraphics[width=\linewidth]{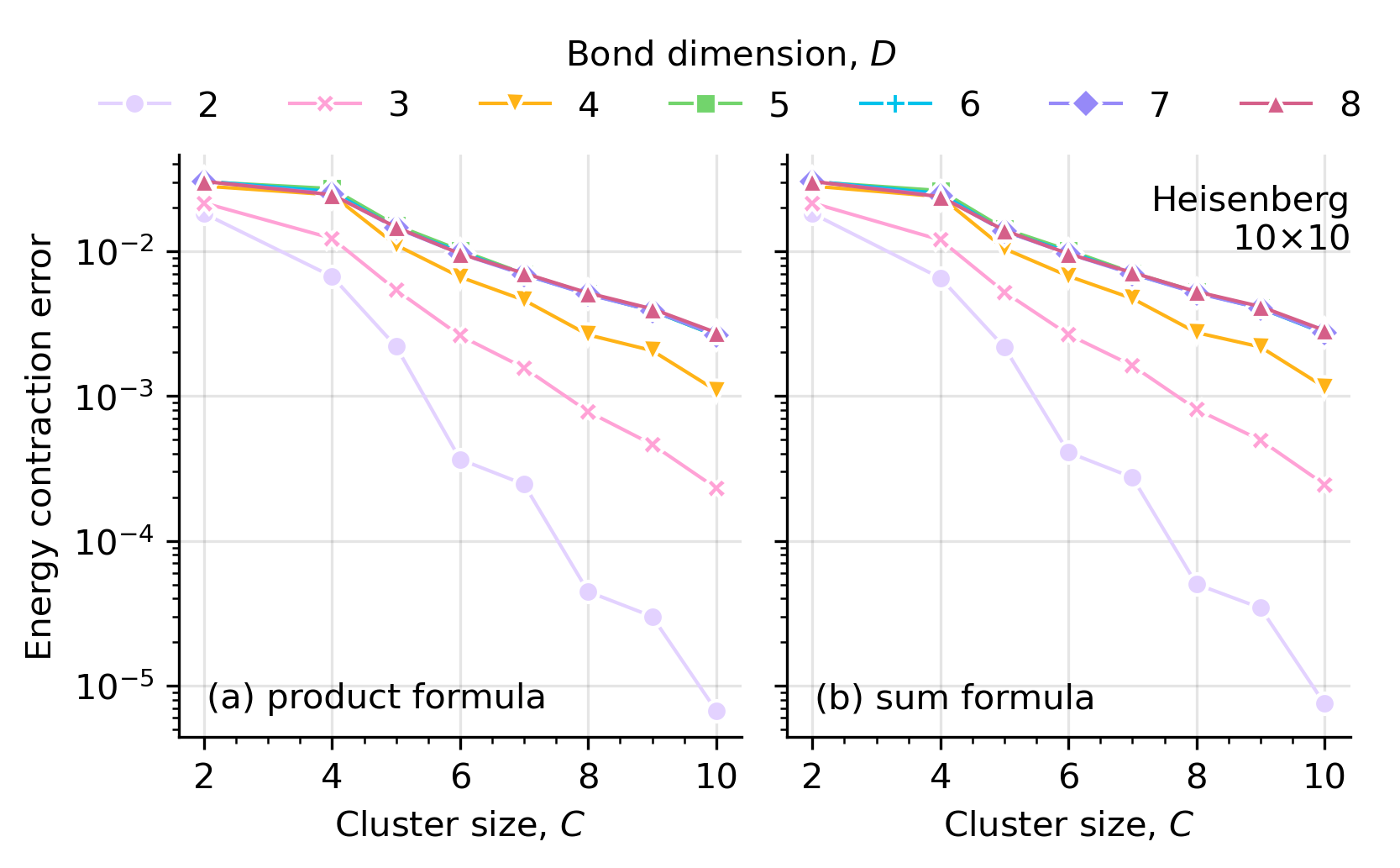}
    \caption{
    Relative energy contraction error for the 2D OBC Heisenberg model with SU-optimized PEPS, as a function of cluster size $C$ and bond dimension $D$, computed with the loop cluster expansion (without extrapolation).
    (a) error using product formula, Eq.~\eqref{eq:product_cluster_expansion}.
    (b) error using sum formula, Eq.~\eqref{eq:sum_cluster_expansion}.
    }
    \label{fig:energy-contraction-error-square-obc-heis}
\end{figure}

\begin{figure}[t!]
    \centering
    \includegraphics[width=\linewidth]{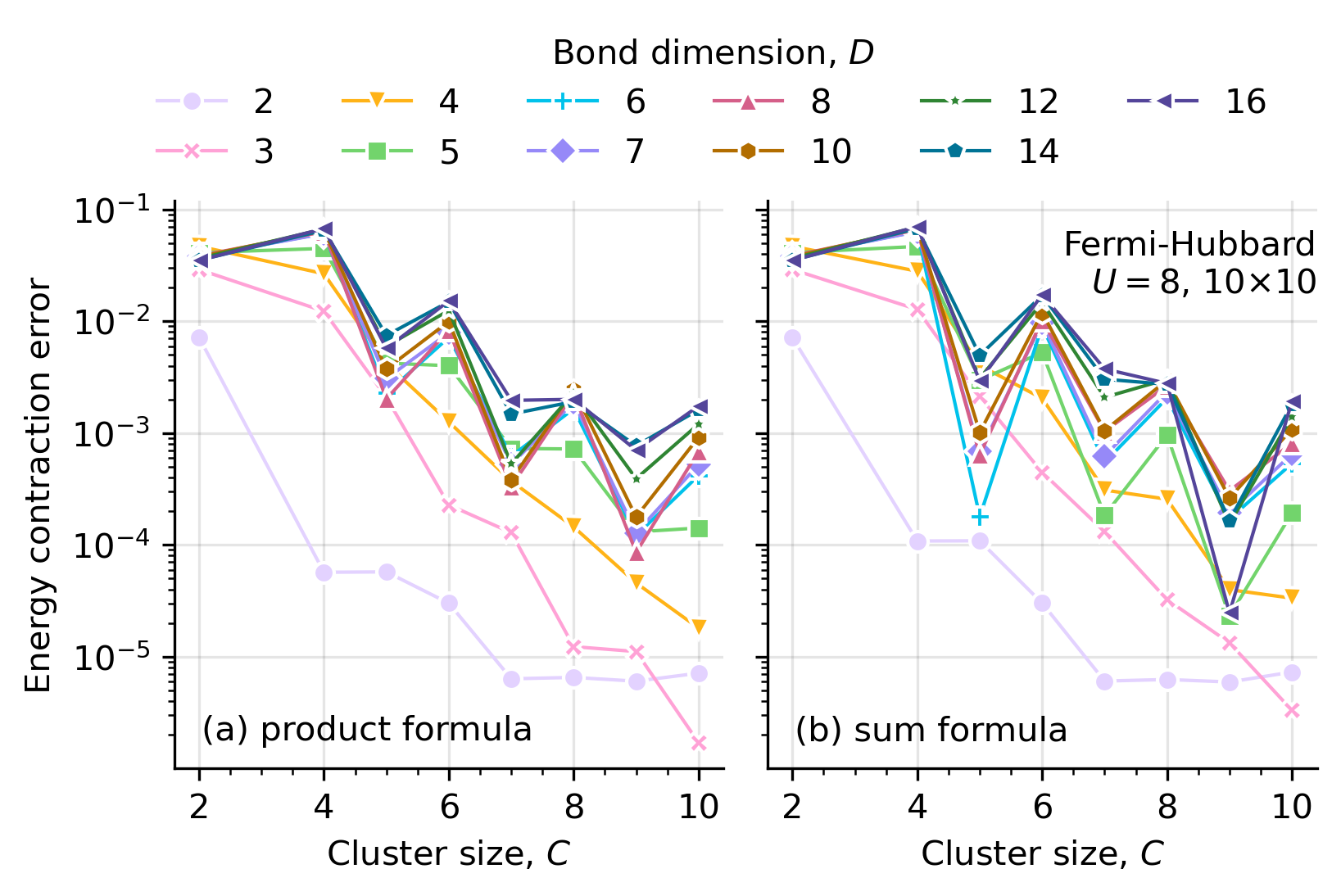}
    \caption{
    Relative energy contraction error for the 2D OBC Fermi-Hubbard mode, $U=8$ at half-filling, with SU-optimized PEPS, as a function of cluster size $C$ and bond dimension $D$, computed with the loop cluster expansion (without extrapolation).
    (a) error using product formula, Eq.~\eqref{eq:product_cluster_expansion}.
    (b) error using sum formula, Eq.~\eqref{eq:sum_cluster_expansion}.
    }
    \label{fig:energy-contraction-error-square-obc-fmhb}
\end{figure}

In the 2D square lattice with OBC, we can perform \emph{boundary contractions} -- sweeping an MPS over the full lattice~\cite{verstraete2004renormalization,Jordaorus2008} -- to get very accurate estimates of the PEPS energy as a reference, and thus, we can quantify the contraction error from the loop cluster expansions.
In all cases, numerically exact ground-state energies are in principle available via quantum Monte Carlo (QMC)~\cite{todo2001cluster,motta2018ab}. 

We prepare each PEPS at bond dimension $D$ by starting with the state from $D{-}1$ and evolving with imaginary time step $\tau {=} 0.5 D^{-3/2}$ until gauges $\Lambda_{ij}$ equilibrate.
We then equilibrate the gauges without any gates (equivalent to BP) to reach a fixed point.
We employ either $Z_2$ or $U(1)$ Abelian symmetry~\cite{Singh2010, Singh2011} to improve efficiency and access larger bond dimensions, and for the fermionic problems, we employ the local fermionic tensor approach~\cite{guGrassmannTensorNetwork2010,mortierFermionicTensorNetwork2025,gaoFermionicTensorNetwork2025}. We do not assume any spatial symmetry and treat the systems as finite and inhomogeneous.

In Fig.~\ref{fig:energy-cluster-vs-expansion}, we show convergence of the loop cluster expansion as a function of cluster size $C$ for computing the ground-state energy from two PEPS on the Heisenberg and FH models defined on a $10{\times}10$ square OBC lattice, compared against the `single cluster' method and reference boundary contraction. 
The single clusters are chosen as the union of all sub-loops up to a corresponding size, for example, the size 12 single cluster is the union of all sub-loops of size 5 as in Fig.~\ref{fig:schematic}.
The loop cluster expansion converges significantly faster than the single-cluster method, requiring roughly half the cluster size to achieve similar accuracy. This is because the dominant contributions from a single large cluster already come from smaller individual clusters within it.

In Figs.~\ref{fig:energy-contraction-error-square-obc-tfim}, ~\ref{fig:energy-contraction-error-square-obc-heis}, and ~\ref{fig:energy-contraction-error-square-obc-fmhb}, we show the relative energy contraction error of the loop cluster expansion as a function of cluster size $C$ and bond dimension $D$ for the TFIM, Heisenberg, and FH models, respectively, including both the product and sum formulas.
The reference boundary contraction energies are computed using a bond dimension of $\chi{{=}}256$ and are converged to $\ll 10^{-6}$ accuracy.  
The loop cluster expansion convergence is generally faster at smaller bond dimensions, but beyond a certain bond dimension, the rate of convergence appears to be insensitive to bond dimension.
In all three models, we see an approximately exponential suppression of error with cluster size $C$, though this trend is strictly monotonic only in the Heisenberg model (Fig.~\ref{fig:energy-contraction-error-square-obc-heis}) and at smaller $D$ in the other two models.
The interplay between model, simple update wavefunction, and specific cluster shapes introduced at each $C$ likely accounts for the more oscillatory convergence in the TFIM and FH models.
For the TFIM (Fig.~\ref{fig:energy-contraction-error-square-obc-tfim}) and Heisenberg model (Fig.~\ref{fig:energy-contraction-error-square-obc-heis}), the product and sum formulas produce essentially identical results. 
For the Fermi-Hubbard model (Fig.~\ref{fig:energy-contraction-error-square-obc-fmhb}), the sum formula has a slightly larger error (other than at $C=5$). For the remaining results, we use the product formula.

\begin{figure}[t!]
    \centering
    \includegraphics[width=1.0\linewidth]{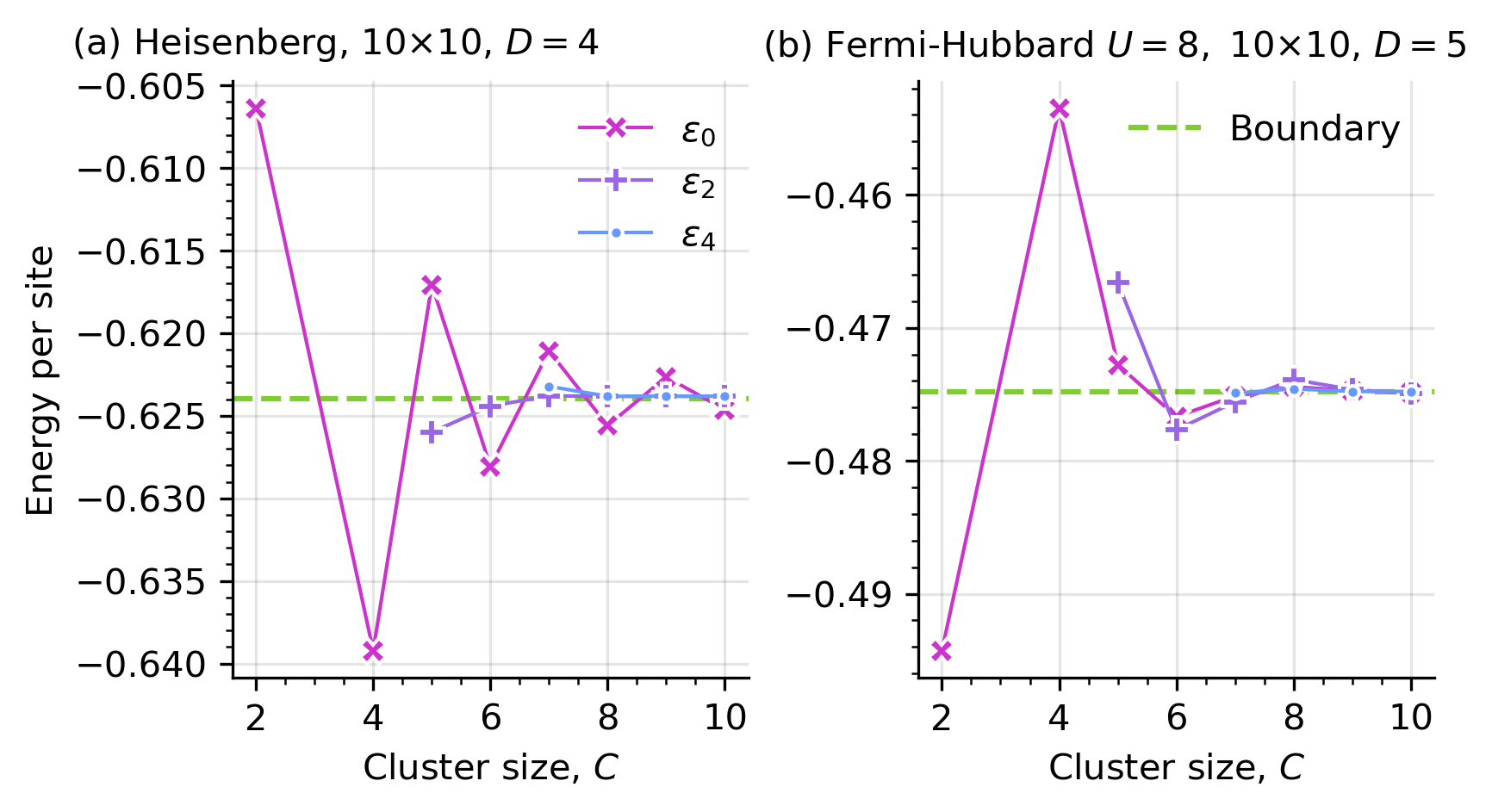}
    \caption{
    Examples of Wynn extrapolation for the same two examples as Fig.~\ref{fig:energy-cluster-vs-expansion}.
    $\epsilon_0$, $\epsilon_2$ and $\epsilon_4$ are the zeroth, 2nd order and 4th order sequences. We use the final value of $\epsilon_4$ as our extrapolated value, and the average final gradient (see main text) across $\epsilon_0$, $\epsilon_2$, and $\epsilon_4$ as an estimate of the error bar.
    }
    \label{fig:error-wynn-extrapolation}
\end{figure}

To obtain reliable estimates in the infinite-cluster limit, we can employ an appropriate extrapolation scheme. Owing to the non-monotonic convergence behavior observed here, we adopt Wynn’s epsilon algorithm~\cite{wynnDeviceComputingEmSn1956}, a sequence acceleration method in common use within numerical linked-cluster expansions~\cite{rigolNumericalLinkedClusterApproach2006,rigolNumericalLinkedclusterAlgorithms2007,rigolNumericalLinkedclusterAlgorithms2007a,tangShortIntroductionNumerical2013a}.
Taking the energy per site, $E_C$, estimated with increasing $C$ to be a converging sequence, we define transformed sequences $\epsilon_{-1}(E_C)=0$, $\epsilon_{0}(E_C)=E_C$ and 
\begin{equation}
\epsilon_{k + 1}(E_C) = \epsilon_{k - 1}(E_{C+1}) + \dfrac{1}{\epsilon_{k}(E_{C+1}) - \epsilon_{k}(E_C)}~.
\label{eq:wynn-series}
\end{equation}
Each transformation yields a shorter and generally smoother sequence.
Only even $k$ sequences give good approximations of the sequence limit, and these are equivalent to diagonal $[k / 2, k / 2]$ Pad\'e approximants.
Examples of this transformation are shown in Fig.~\ref{fig:error-wynn-extrapolation}(a) and (b) for the same data as Fig.~\ref{fig:energy-cluster-vs-expansion}.
We take the $k{{=}}4$ sequence at the largest available cluster size $C_\mathrm{max}$ as our final extrapolated value $E{=}\epsilon_4(C_\mathrm{max})$. Across all 2D OBC data where we can access contraction errors, TFIM with $B_X{=}-3$ ($L{=}4,6,8,10$, $D=2\ldots8$), Heisenberg ($L{=}4,6,8,10$, $D=2\ldots8$) and FH ($L{=}4,6,8,10$, $D=2\ldots16$) we find this extrapolation reduces the median error by a factor of ${\approx} 8 \times$.
Empirically we also find that the average final gradient 
$
\delta E {=} (
|\Delta \epsilon_0| +
|\Delta \epsilon_2| +
|\Delta \epsilon_4| ) / 3
$
where $\Delta \epsilon_k = \epsilon_k(E_{C_\mathrm{max}}) - \epsilon_k(E_{C_\mathrm{max} - 1})$
gives a relatively conservative estimate of the real contraction error. We use this as an approximate error bar on our extrapolated values.

\begin{figure}[t!]
    \centering
    \includegraphics[width=1.0\linewidth]{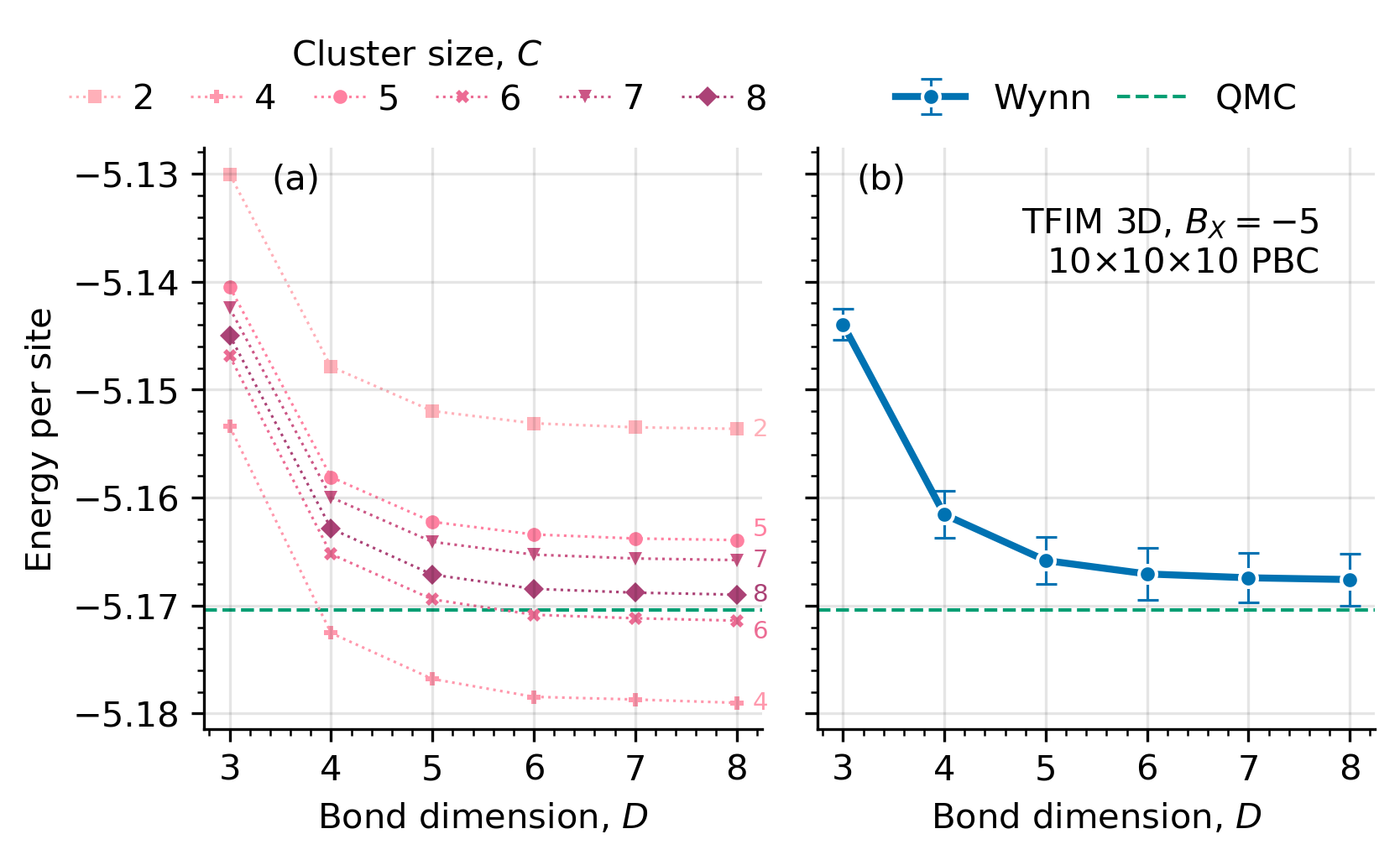}
    \caption{
    Energy per site for the 3D PBC TFIM with SU-optimized PEPS as a function of bond dimension $D$, computed using the loop cluster expansion. (a)~Raw estimates at each cluster size $C$. (b)~Wynn-extrapolated values $E$ and error bars $\delta E$. Dashed green line: reference QMC result~\cite{todo2001cluster,albuquerque2007alps,bauer2011alps}.
    \label{fig:models-energy-vs-D-tfim}
    }
\end{figure}

\begin{figure}[t!]
    \centering
    \includegraphics[width=1.0\linewidth]{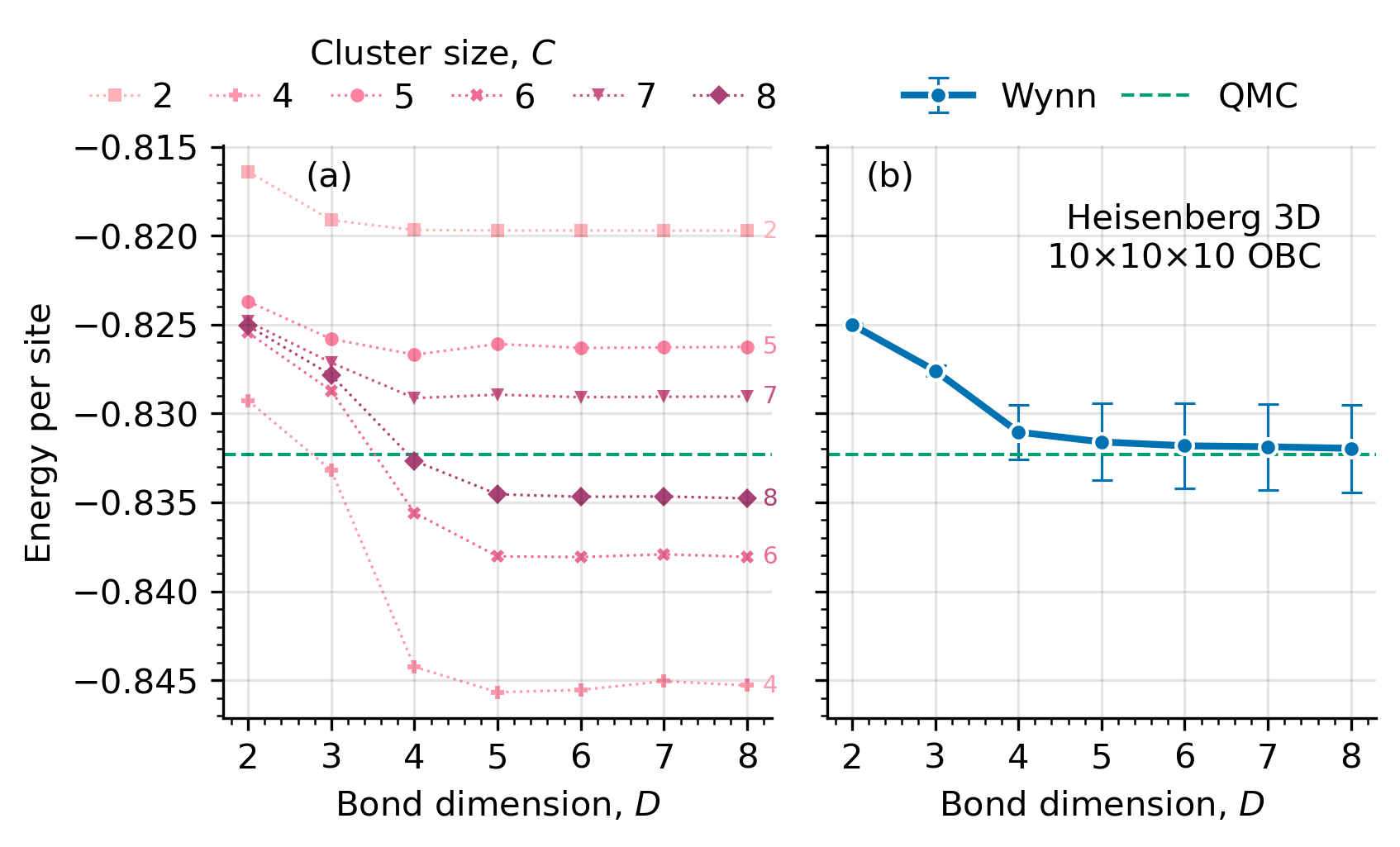}
    \caption{
    Energy per site for the 3D OBC Heisenberg model with SU-optimized PEPS as a function of bond dimension $D$, computed using the loop cluster expansion. (a)~Raw estimates at each cluster size $C$. (b)~Wynn-extrapolated values $E$ and error bars $\delta E$. Dashed green line: reference QMC result~\cite{todo2001cluster,albuquerque2007alps,bauer2011alps}.
    \label{fig:models-energy-vs-D-heis}
    }
\end{figure}

\begin{figure}[t!]
    \centering
    \includegraphics[width=1.0\linewidth]{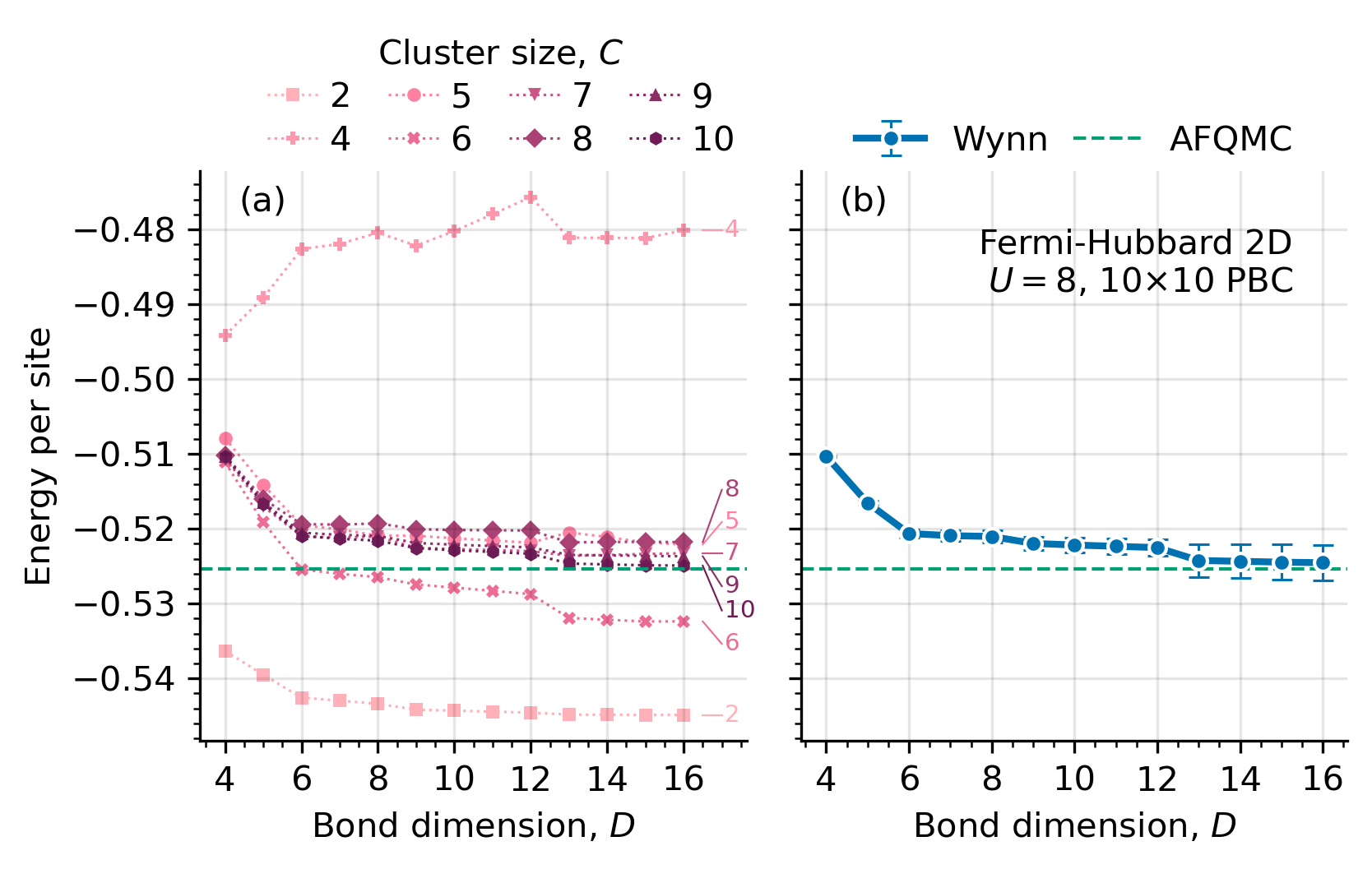}
    \caption{
    Energy per site for the 2D PBC Fermi-Hubbard model, $U{=}8$ at half-filling, with SU-optimized PEPS as a function of bond dimension $D$, computed using the loop cluster expansion. (a)~Raw estimates at each cluster size $C$. (b)~Wynn-extrapolated values $E$ and error bars $\delta E$. Dashed green line: reference AFQMC result~\cite{motta2018ab,mahajan2023response,ad_afqmc}.
    \label{fig:models-energy-vs-D-fmhb2d}
    }
\end{figure}

\begin{figure}[t!]
    \centering
    \includegraphics[width=1.0\linewidth]{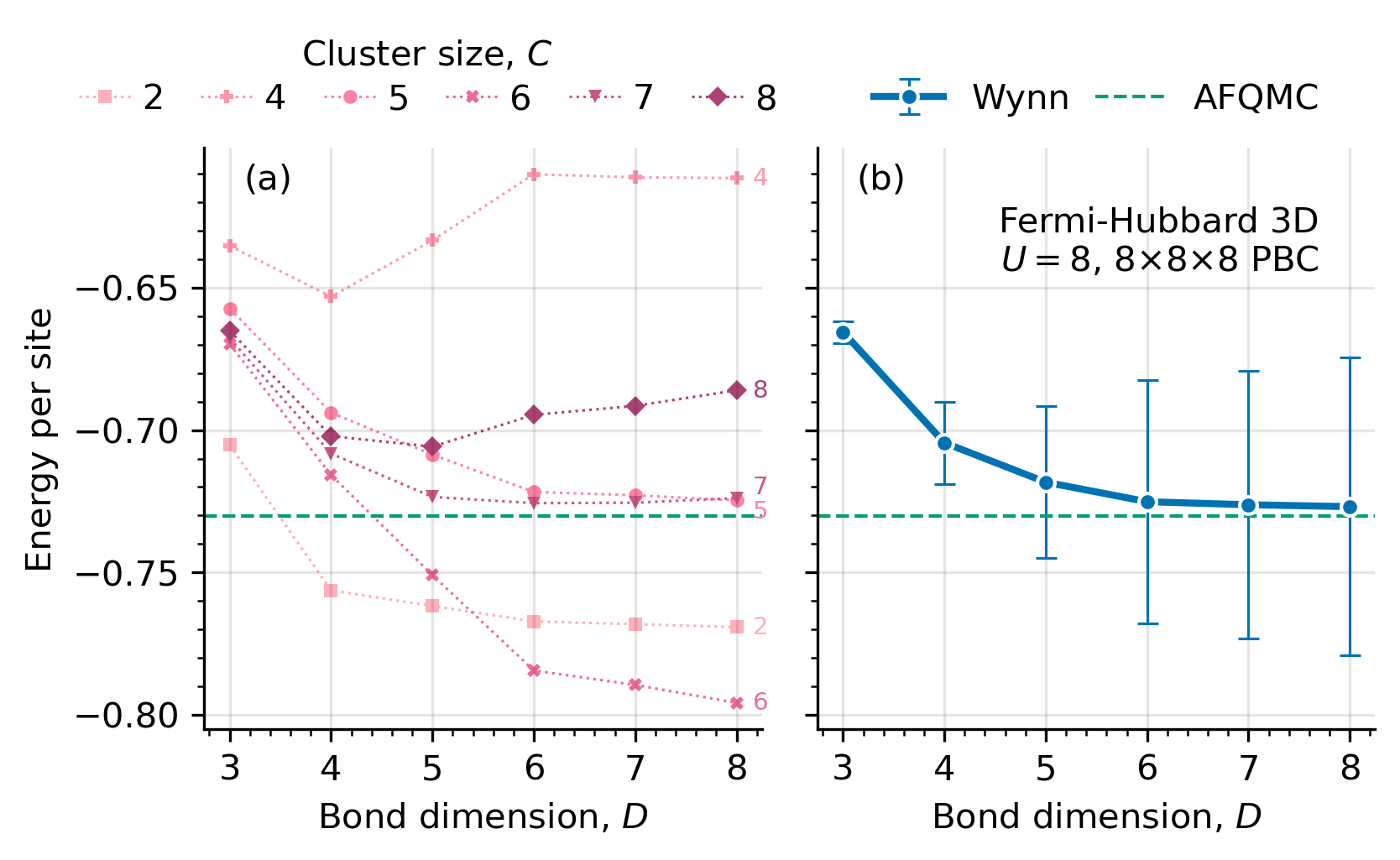}
    \caption{
    Energy per site for the 3D PBC Fermi-Hubbard model, $U{=}8$ at half-filling, with SU-optimized PEPS as a function of bond dimension $D$, computed using the loop cluster expansion. (a)~Raw estimates at each cluster size $C$. (b)~Wynn-extrapolated values $E$ and error bars $\delta E$. Dashed green line: reference AFQMC result~\cite{motta2018ab,mahajan2023response,ad_afqmc}.
    \label{fig:models-energy-vs-D-fmhb3d}
    }
\end{figure}

\begin{table*}[t!]
\begin{tabular}{ |c|c|c|c|c| } 
 \hline
 \makecell{Model} & 
 \makecell{$C_\mathrm{max}$} & 
 \makecell{$D_\mathrm{max}$} & 
 \makecell{$E(\delta E)$} & 
 \makecell{$E_\mathrm{QMC}$} \\
  \hline
  \makecell{
 TFIM 3D, $B_X{{=}}-5$, $10{{\times}}10{{\times}}10$ PBC
 } & 8 & 8 & $-5.1676(24)$ & $-5.170442(95)$ \\
 \hline
 \makecell{
 Heisenberg 3D, $10{{\times}}10{{\times}}10$ OBC 
 } & 8 & 8 & $-0.8320(25)$ & $-0.832311(14)$ \\
 \hline
 \makecell{
 Fermi-Hubbard 2D, $U=8$, $10{{\times}}10$ PBC
 } & 10 & 16 & $-0.5245(23)$ &  $-0.52540(30)$ \\
 \hline
 \makecell{
 Fermi-Hubbard 3D, $U=8$, $8{{\times}}8{{\times}}8$ PBC
 } & 8 & 8 & $-0.727(52)$ & $-0.730261(22)$ \\
 \hline
\end{tabular}
\caption{Final energies (errors) per site, $E(\delta E)$, for PEPS ground-states optimized using SU, computed with the loop cluster expansion and Wynn extrapolation, compared to reference QMC results, $E_\mathrm{QMC}$, computed either using the stochastic series expansion loop algorithm~\cite{todo2001cluster,albuquerque2007alps,bauer2011alps} for the TFIM and Heisenberg model or auxiliary field QMC (AFQMC) for the FH model~\cite{motta2018ab,mahajan2023response,ad_afqmc}.
$C_\mathrm{max}$ denotes the largest cluster size, and $D_\mathrm{max}$ the largest bond dimension, used to compute $E$.
}
\label{tab:results}
\end{table*}

Finally, we use the energy, $E$, and error, $\delta E$, estimated using Wynn extrapolation, to study the performance of SU optimization. Here, the use of the loop cluster expansion allows us to assess this variational error at large bond dimensions and in geometries/boundary conditions where other contraction methods are not readily applied. 
We choose the examples below to use a practical amount of computation, e.g. $\sim$~a few days on a 8-core CPU for the most expensive points.
In Fig.~\ref{fig:models-energy-vs-D-tfim}, we show $E$ as a function of $D$ for the TFIM with $B_X{{=}}-5$ on a 3D $10{{\times}}10{{\times}}10$ lattice with PBC.
For reference we also show the raw cluster data used to perform the extrapolation.
In Fig.~\ref{fig:models-energy-vs-D-heis}, we study the Heisenberg model on a 3D $10{{\times}}10{{\times}}10$ lattice with OBC.
In Fig.~\ref{fig:models-energy-vs-D-fmhb2d}, we move to the Fermi-Hubbard model at half-filling with $U{{=}}8$ on a 2D $10{{\times}}10$ lattice with PBC.
Finally, in Fig.~\ref{fig:models-energy-vs-D-fmhb3d} we study the same model 
on a 3D $8{{\times}}8{{\times}}8$ lattice with PBC.
The larger bandwidth in 3D ($12t$ in units of hopping $t$, versus $8t$ in 2D) means that the fermions are more itinerant in 3D than in 2D. We find that this system poses the largest challenge for the loop cluster expansion method, with convergence with $C$ (unlike the TFIM and Heisenberg case) significantly slower than in 2D, as reflected in $\delta E$. 
In Table~\ref{tab:results}, we collate the maximum cluster size and bond dimension used for each model, and compare the final energy and error with reference QMC calculations~\cite{bauer2011alps,motta2018ab,qin2016benchmark}.

%%%%%%%%%%%%%%%%%%%%%%%%%%%%%%%%%%%%%%%%%%%%%%%%%%%%%%%%%%%%%%%%

\section{Conclusions}

In this work, we described and analyzed the loop cluster expansion of tensor networks introduced in~\cite{park2025simulatingquantumdynamicstwodimensional}.
We focused on the estimation of local observable expectation values and used them to evaluate the ground-state energies of PEPS of a range of physical models. For 2D models where we had numerically converged data from boundary contraction methods, we observed approximately exponential convergence of the contraction error for the energy with cluster size.
We also exploited Wynn's epsilon algorithm for extrapolations to the infinite-cluster limit.
We showcased the practical applicability of the method to tensor network contraction in more complicated geometries, such as with periodic boundary conditions or for 3D systems, where applying conventional tensor network contraction methods becomes challenging.  

We expect that the loop cluster expansion analyzed in this work has potential beyond computing local observables. For example, the loop cluster expansion can be used to approximate the environment when compressing tensors obtained from real and imaginary time-evolution, generalizing the so-called `cluster update'~\cite{Lubasch_2014, Lubasch2014prb}. The logic of the cluster expansion can also be applied to design a new message-passing routine, in the spirit of generalized belief propagation. Finally, it is natural to extend the size of clusters entering into the loop cluster expansion through the use of traditional approximate tensor network contraction of the largest clusters.

\begin{acknowledgments}

The tensor network library \texttt{quimb}~\cite{gray2018quimb} was used with the Abelian symmetric and fermionic array backend \texttt{symmray},~\cite{gaoFermionicTensorNetwork2025,symmray}, with \texttt{cotengra}~\cite{gray2021hyper} for cluster contractions in the numerical experiments.
The authors thank Ashley Milsted for helpful discussions. This work was supported by
the US Department of Energy, Office of Science, Accelerated Research in Quantum Computing Centers, Quantum Utility through Advanced Computational Quantum Algorithms, through Award No. DE-SC0025572.

\end{acknowledgments}

\vspace{1em}
\emph{Note.--} During the submission of this manuscript, a related preprint~\cite{midha2025beliefpropagationclustercorrectedtensor} appeared. Ref.~\cite{midha2025beliefpropagationclustercorrectedtensor} builds upon the loop series expansion of the excitation space~\cite{ChertkovChernyak2006, ChertkovChernyak2006b, evenbly2025loopseriesexpansionstensor}. However, instead of expanding the partition function, Ref.~\cite{midha2025beliefpropagationclustercorrectedtensor} expands the free energy using clusters of connected loops. In contrast, our work computes the loopy clusters and utilizes them in a manner analogous to the linked cluster expansion.

%%%%%%%%%%%%%%%%%%%%%%%%%%%%%%%%%%%%%%%%%%%%%%%%%%%%%%%%%%%%%%%%

\bibliography{main}

@article{pancotti2023one,
  title={One-step replica symmetry breaking in the language of tensor networks},
  author={Pancotti, Nicola and Gray, Johnnie},
  journal={arXiv preprint arXiv:2306.15004},
  year={2023},
  url = {https://arxiv.org/abs/2306.15004}
}

@article{gray2024hyperoptimized,
  title={Hyperoptimized approximate contraction of tensor networks with arbitrary geometry},
  author={Gray, Johnnie and Chan, Garnet Kin-Lic},
  journal={Physical Review X},
  volume={14},
  number={1},
  pages={011009},
  year={2024},
  publisher={American Physical Society},
  doi = {10.1103/PhysRevX.14.011009},
  url = {https://link.aps.org/doi/10.1103/PhysRevX.14.011009}
}

@article{SCHOLLWOCK201196,
title = {The density-matrix renormalization group in the age of matrix product states},
journal = {Annals of Physics},
volume = {326},
number = {1},
pages = {96-192},
year = {2011},
note = {January 2011 Special Issue},
issn = {0003-4916},
doi = {https://doi.org/10.1016/j.aop.2010.09.012},
url = {https://www.sciencedirect.com/science/article/pii/S0003491610001752},
author = {Ulrich Schollw\"ock},
}

@article{CiracPerezSchuch2021,
  title = {Matrix product states and projected entangled pair states: Concepts, symmetries, theorems},
  author = {Cirac, J. Ignacio and P\'erez-Garc\'{\i}a, David and Schuch, Norbert and Verstraete, Frank},
  journal = {Rev. Mod. Phys.},
  volume = {93},
  issue = {4},
  pages = {045003},
  numpages = {65},
  year = {2021},
  month = {Dec},
  publisher = {American Physical Society},
  doi = {10.1103/RevModPhys.93.045003},
  url = {https://link.aps.org/doi/10.1103/RevModPhys.93.045003}
}

@article{white1992density,
  title = {Density matrix formulation for quantum renormalization groups},
  author = {White, Steven R.},
  journal = {Phys. Rev. Lett.},
  volume = {69},
  issue = {19},
  pages = {2863--2866},
  numpages = {0},
  year = {1992},
  month = {Nov},
  publisher = {American Physical Society},
  doi = {10.1103/PhysRevLett.69.2863},
  url = {https://link.aps.org/doi/10.1103/PhysRevLett.69.2863}
}

@article{white1993density,
  title = {Density-matrix algorithms for quantum renormalization groups},
  author = {White, Steven R.},
  journal = {Phys. Rev. B},
  volume = {48},
  issue = {14},
  pages = {10345--10356},
  numpages = {0},
  year = {1993},
  month = {Oct},
  publisher = {American Physical Society},
  doi = {10.1103/PhysRevB.48.10345},
  url = {https://link.aps.org/doi/10.1103/PhysRevB.48.10345}
}

@article{verstraete2004renormalization,
  title={Renormalization algorithms for quantum-many body systems in two and higher dimensions},
  author={Verstraete, Frank and Cirac, J Ignacio},
  journal={arXiv preprint cond-mat/0407066},
  year={2004},
  url = {https://arxiv.org/abs/cond-mat/0407066}
}

@article{jiangAccurateDeterminationTensor2008a,
  title = {Accurate {{Determination}} of {{Tensor Network State}} of {{Quantum Lattice Models}} in {{Two Dimensions}}},
  author = {Jiang, H. C. and Weng, Z. Y. and Xiang, T.},
  year = {2008},
  month = aug,
  journal = {Physical Review Letters},
  volume = {101},
  number = {9},
  pages = {090603},
  publisher = {American Physical Society},
  doi = {10.1103/PhysRevLett.101.090603},
  urldate = {2025-10-06},
  url = {https://link.aps.org/doi/10.1103/PhysRevLett.101.090603}
}

@misc{guGrassmannTensorNetwork2010,
  title = {Grassmann Tensor Network States and Its Renormalization for Strongly Correlated Fermionic and Bosonic States},
  author = {Gu, Zheng-Cheng and Verstraete, Frank and Wen, Xiao-Gang},
  year = {2010},
  month = apr,
  number = {arXiv:1004.2563},
  eprint = {1004.2563},
  primaryclass = {cond-mat, physics:quant-ph},
  publisher = {arXiv},
  doi = {10.48550/arXiv.1004.2563},
  urldate = {2024-05-23},
  archiveprefix = {arXiv},
  url = {https://arxiv.org/abs/1004.2563}
}

@article{mortierFermionicTensorNetwork2025,
  title = {Fermionic Tensor Network Methods},
  author = {Mortier, Quinten and Devos, Lukas and Burgelman, Lander and Vanhecke, Bram and Bultinck, Nick and Verstraete, Frank and Haegeman, Jutho and Vanderstraeten, Laurens},
  year = {2025},
  month = jan,
  journal = {SciPost Physics},
  volume = {18},
  number = {1},
  pages = {012},
  issn = {2542-4653},
  doi = {10.21468/SciPostPhys.18.1.012},
  urldate = {2025-10-06},
  langid = {english},
  url = {https://scipost.org/10.21468/SciPostPhys.18.1.012}
}

@article{gaoFermionicTensorNetwork2025,
  title = {Fermionic Tensor Network Contraction for Arbitrary Geometries},
  author = {Gao, Yang and Zhai, Huanchen and Gray, Johnnie and Peng, Ruojing and Park, Gunhee and Liu, Wen-Yuan and Kj{\o}nstad, Eirik F. and Chan, Garnet Kin-Lic},
  year = {2025},
  month = may,
  journal = {Physical Review Research},
  volume = {7},
  number = {2},
  pages = {023193},
  publisher = {American Physical Society},
  doi = {10.1103/PhysRevResearch.7.023193},
  urldate = {2025-10-06},
  url = {https://link.aps.org/doi/10.1103/PhysRevResearch.7.023193}
}

@article{wynnDeviceComputingEmSn1956,
  title = {On a {{Device}} for {{Computing}} the Em({{Sn}}) {{Transformation}}},
  author = {Wynn, P.},
  year = {1956},
  journal = {Mathematical Tables and Other Aids to Computation},
  volume = {10},
  number = {54},
  eprint = {2002183},
  eprinttype = {jstor},
  pages = {91--96},
  publisher = {American Mathematical Society},
  issn = {0891-6837},
  doi = {10.2307/2002183},
  urldate = {2025-10-06},
  url = {https://doi.org/10.2307/2002183}
}

@article{evenbly2025loopseriesexpansionstensor,
  title = {Loop series expansions for tensor networks},
  author = {Evenbly, Glen and Pancotti, Nicola and Milsted, Ashley and Gray, Johnnie and Chan, Garnet Kin-Lic},
  journal = {Phys. Rev. Res.},
  volume = {8},
  issue = {1},
  pages = {013245},
  numpages = {14},
  year = {2026},
  month = {Mar},
  publisher = {American Physical Society},
  doi = {10.1103/vqks-cr6x},
  url = {https://link.aps.org/doi/10.1103/vqks-cr6x}
}

@article{rigolNumericalLinkedclusterAlgorithms2007,
  title = {Numerical Linked-Cluster Algorithms. {{I}}. {{Spin}} Systems on Square, Triangular, and Kagom\'e Lattices},
  author = {Rigol, Marcos and Bryant, Tyler and Singh, Rajiv R. P.},
  year = {2007},
  month = jun,
  journal = {Physical Review E},
  volume = {75},
  number = {6},
  pages = {061118},
  publisher = {American Physical Society},
  doi = {10.1103/PhysRevE.75.061118},
  urldate = {2025-10-06},
  url = {https://link.aps.org/doi/10.1103/PhysRevE.75.061118}
}

@article{rigolNumericalLinkedclusterAlgorithms2007a,
  title = {Numerical Linked-Cluster Algorithms. {{II}}. $t{{-}}J$ Models on the Square Lattice},
  author = {Rigol, Marcos and Bryant, Tyler and Singh, Rajiv R. P.},
  year = {2007},
  month = jun,
  journal = {Physical Review E},
  volume = {75},
  number = {6},
  pages = {061119},
  publisher = {American Physical Society},
  doi = {10.1103/PhysRevE.75.061119},
  urldate = {2025-10-06},
  url = {https://link.aps.org/doi/10.1103/PhysRevE.75.061119}
}

@article{rigolNumericalLinkedClusterApproach2006,
  title = {Numerical {{Linked-Cluster Approach}} to {{Quantum Lattice Models}}},
  author = {Rigol, Marcos and Bryant, Tyler and Singh, Rajiv R. P.},
  year = {2006},
  month = nov,
  journal = {Physical Review Letters},
  volume = {97},
  number = {18},
  pages = {187202},
  publisher = {American Physical Society},
  doi = {10.1103/PhysRevLett.97.187202},
  urldate = {2025-10-06},
  url = {https://link.aps.org/doi/10.1103/PhysRevLett.97.187202}
}

@article{tangShortIntroductionNumerical2013a,
  title = {A Short Introduction to Numerical Linked-Cluster Expansions},
  author = {Tang, Baoming and Khatami, Ehsan and Rigol, Marcos},
  year = {2013},
  month = mar,
  journal = {Computer Physics Communications},
  volume = {184},
  number = {3},
  pages = {557--564},
  issn = {0010-4655},
  doi = {10.1016/j.cpc.2012.10.008},
  urldate = {2025-10-06},
  url = {https://doi.org/10.1016/j.cpc.2012.10.008}
}

@inproceedings{welling2012,
author = {Welling, Max and Gelfand, Andrew E. and Ihler, Alexander},
title = {A cluster-cumulant expansion at the fixed points of belief propagation},
year = {2012},
isbn = {9780974903989},
publisher = {AUAI Press},
address = {Arlington, Virginia, USA},
booktitle = {Proceedings of the Twenty-Eighth Conference on Uncertainty in Artificial Intelligence},
pages = {883–892},
numpages = {10},
location = {Catalina Island, CA},
series = {UAI'12},
  url = {https://arxiv.org/abs/1210.4916}
}

@article{park2025simulatingquantumdynamicstwodimensional,
  title = {Simulating quantum dynamics in two-dimensional lattices with tensor network influence functional belief propagation},
  author = {Park, Gunhee and Gray, Johnnie and Chan, Garnet Kin-Lic},
  journal = {Phys. Rev. B},
  volume = {112},
  issue = {17},
  pages = {174310},
  numpages = {16},
  year = {2025},
  month = {Nov},
  publisher = {American Physical Society},
  doi = {10.1103/7jzt-xhn6},
  url = {https://link.aps.org/doi/10.1103/7jzt-xhn6}
}

@article{Vidal2003,
  title = {Efficient Classical Simulation of Slightly Entangled Quantum Computations},
  author = {Vidal, Guifr\'e},
  journal = {Phys. Rev. Lett.},
  volume = {91},
  issue = {14},
  pages = {147902},
  numpages = {4},
  year = {2003},
  month = {Oct},
  publisher = {American Physical Society},
  doi = {10.1103/PhysRevLett.91.147902},
  url = {https://link.aps.org/doi/10.1103/PhysRevLett.91.147902}
}

@article{Vidal2004,
  title = {Efficient Simulation of One-Dimensional Quantum Many-Body Systems},
  author = {Vidal, Guifr\'e},
  journal = {Phys. Rev. Lett.},
  volume = {93},
  issue = {4},
  pages = {040502},
  numpages = {4},
  year = {2004},
  month = {Jul},
  publisher = {American Physical Society},
  doi = {10.1103/PhysRevLett.93.040502},
  url = {https://link.aps.org/doi/10.1103/PhysRevLett.93.040502}
}

@article{Vidal2007,
  title = {Classical Simulation of Infinite-Size Quantum Lattice Systems in One Spatial Dimension},
  author = {Vidal, G.},
  journal = {Phys. Rev. Lett.},
  volume = {98},
  issue = {7},
  pages = {070201},
  numpages = {4},
  year = {2007},
  month = {Feb},
  publisher = {American Physical Society},
  doi = {10.1103/PhysRevLett.98.070201},
  url = {https://link.aps.org/doi/10.1103/PhysRevLett.98.070201}
}

@Article{Tindall2023Scipost,
	title={{Gauging tensor networks with belief propagation}},
	author={Joseph Tindall and Matt Fishman},
	journal={SciPost Phys.},
	volume={15},
	pages={222},
	year={2023},
	publisher={SciPost},
	doi={10.21468/SciPostPhys.15.6.222},
	url={https://scipost.org/10.21468/SciPostPhys.15.6.222},
}

@article{AlkabetzArad2021,
  title = {Tensor networks contraction and the belief propagation algorithm},
  author = {Alkabetz, R. and Arad, I.},
  journal = {Phys. Rev. Res.},
  volume = {3},
  issue = {2},
  pages = {023073},
  numpages = {12},
  year = {2021},
  month = {Apr},
  publisher = {American Physical Society},
  doi = {10.1103/PhysRevResearch.3.023073},
  url = {https://link.aps.org/doi/10.1103/PhysRevResearch.3.023073}
}

@article{Lubasch_2014,
doi = {10.1088/1367-2630/16/3/033014},
url = {https://doi.org/10.1088/1367-2630/16/3/033014},
year = {2014},
month = {mar},
publisher = {IOP Publishing},
volume = {16},
number = {3},
pages = {033014},
author = {Lubasch, Michael and Cirac, J Ignacio and Bañuls, Mari-Carmen},
title = {Unifying projected entangled pair state contractions},
journal = {New Journal of Physics},
}

@article{Jahromi2019,
  title = {Universal Tensor-Network Algorithm for Any Infinite Lattice},
  author = {Jahromi, Saeed S. and Or{\'u}s, Rom{\'a}n},
  year = {2019},
  month = may,
  journal = {Physical Review B},
  volume = {99},
  number = {19},
  pages = {195105},
  publisher = {American Physical Society},
  doi = {10.1103/PhysRevB.99.195105},
  urldate = {2021-02-23},
  url = {https://link.aps.org/doi/10.1103/PhysRevB.99.195105}
}

@article{jahromi2020thermal,
  title={Thermal bosons in 3d optical lattices via tensor networks},
  author={Jahromi, Saeed S and Or{\'u}s, Rom{\'a}n},
  journal={Scientific Reports},
  volume={10},
  number={1},
  pages={19051},
  year={2020},
  publisher={Nature Publishing Group UK London},
  url={https://doi.org/10.1038/s41598-020-75548-x}
}

@article{Vlaar2021,
  title = {Simulation of three-dimensional quantum systems with projected entangled-pair states},
  author = {Vlaar, Patrick C. G. and Corboz, Philippe},
  journal = {Phys. Rev. B},
  volume = {103},
  issue = {20},
  pages = {205137},
  numpages = {11},
  year = {2021},
  month = {May},
  publisher = {American Physical Society},
  doi = {10.1103/PhysRevB.103.205137},
  url = {https://link.aps.org/doi/10.1103/PhysRevB.103.205137}
}

@inproceedings{yedidiaGeneralizedBeliefPropagation2000,
  title = {Generalized {{Belief Propagation}}},
  booktitle = {Neural {{Information Processing Systems}}},
  author = {Yedidia, J. and Freeman, W. and Weiss, Yair},
  year = {2000},
  urldate = {2023-11-15},
  url = {https://proceedings.neurips.cc/paper/2000/hash/61b1fb3f59e28c67f3925f3c79be81a1-Abstract.html}
}

@incollection{yedidiaUnderstandingBeliefPropagation2003,
  title = {Understanding Belief Propagation and Its Generalizations},
  booktitle = {Exploring Artificial Intelligence in the New Millennium},
  author = {Yedidia, Jonathan S. and Freeman, William T. and Weiss, Yair},
  year = {2003},
  month = jan,
  pages = {239--269},
  publisher = {Morgan Kaufmann Publishers Inc.},
  address = {San Francisco, CA, USA},
  urldate = {2023-11-11},
  isbn = {978-1-55860-811-5},
  url = {https://dl.acm.org/doi/10.5555/779343.779352}
}

@article{yedidiaConstructingFreeenergyApproximations2005,
  title = {Constructing Free-Energy Approximations and Generalized Belief Propagation Algorithms},
  author = {Yedidia, J.S. and Freeman, W.T. and Weiss, Y.},
  year = {2005},
  month = jul,
  journal = {IEEE Transactions on Information Theory},
  volume = {51},
  number = {7},
  pages = {2282--2312},
  issn = {1557-9654},
  doi = {10.1109/TIT.2005.850085},
  urldate = {2023-09-25},
  url = {https://doi.org/10.1109/TIT.2005.850085}
}

@article{Singh2010,
  title = {Tensor network decompositions in the presence of a global symmetry},
  author = {Singh, Sukhwinder and Pfeifer, Robert N. C. and Vidal, Guifr\'e},
  journal = {Phys. Rev. A},
  volume = {82},
  issue = {5},
  pages = {050301},
  numpages = {4},
  year = {2010},
  month = {Nov},
  publisher = {American Physical Society},
  doi = {10.1103/PhysRevA.82.050301},
  url = {https://link.aps.org/doi/10.1103/PhysRevA.82.050301}
}

@article{Singh2011,
  title = {Tensor network states and algorithms in the presence of a global U(1) symmetry},
  author = {Singh, Sukhwinder and Pfeifer, Robert N. C. and Vidal, Guifre},
  journal = {Phys. Rev. B},
  volume = {83},
  issue = {11},
  pages = {115125},
  numpages = {22},
  year = {2011},
  month = {Mar},
  publisher = {American Physical Society},
  doi = {10.1103/PhysRevB.83.115125},
  url = {https://link.aps.org/doi/10.1103/PhysRevB.83.115125}
}

@misc{midha2025beliefpropagationclustercorrectedtensor,
      title={Beyond Belief Propagation: Cluster-Corrected Tensor Network Contraction with Exponential Convergence},
      author={Siddhant Midha and Yifan F. Zhang},
      year={2025},
      eprint={2510.02290},
      archivePrefix={arXiv},
      primaryClass={quant-ph},
      url={https://arxiv.org/abs/2510.02290},
}

@article{Begusic2024,
author = {Tomislav Begu\v{s}i\'{c} and Johnnie Gray  and Garnet Kin-Lic Chan },
title = {Fast and converged classical simulations of evidence for the utility of quantum computing before fault tolerance},
journal = {Science Advances},
volume = {10},
number = {3},
pages = {eadk4321},
year = {2024},
doi = {10.1126/sciadv.adk4321},
URL = {https://www.science.org/doi/abs/10.1126/sciadv.adk4321},
}

@article{Tindall2024ibm,
  title = {Efficient Tensor Network Simulation of IBM's Eagle Kicked Ising Experiment},
  author = {Tindall, Joseph and Fishman, Matthew and Stoudenmire, E. Miles and Sels, Dries},
  journal = {PRX Quantum},
  volume = {5},
  issue = {1},
  pages = {010308},
  numpages = {16},
  year = {2024},
  month = {Jan},
  publisher = {American Physical Society},
  doi = {10.1103/PRXQuantum.5.010308},
  url = {https://link.aps.org/doi/10.1103/PRXQuantum.5.010308}
}

@article{Orus2024ibm,
  title = {Efficient tensor network simulation of IBM's largest quantum processors},
  author = {Patra, Siddhartha and Jahromi, Saeed S. and Singh, Sukhbinder and Or\'us, Rom\'an},
  journal = {Phys. Rev. Res.},
  volume = {6},
  issue = {1},
  pages = {013326},
  numpages = {7},
  year = {2024},
  month = {Mar},
  publisher = {American Physical Society},
  doi = {10.1103/PhysRevResearch.6.013326},
  url = {https://link.aps.org/doi/10.1103/PhysRevResearch.6.013326}
}

@article{Ran2013,
  title = {Theory of network contractor dynamics for exploring thermodynamic properties of two-dimensional quantum lattice models},
  author = {Ran, Shi-Ju and Xi, Bin and Liu, Tao and Su, Gang},
  journal = {Phys. Rev. B},
  volume = {88},
  issue = {6},
  pages = {064407},
  numpages = {8},
  year = {2013},
  month = {Aug},
  publisher = {American Physical Society},
  doi = {10.1103/PhysRevB.88.064407},
  url = {https://link.aps.org/doi/10.1103/PhysRevB.88.064407}
}

@article{RanSuperortho2012,
  title = {Optimized decimation of tensor networks with super-orthogonalization for two-dimensional quantum lattice models},
  author = {Ran, Shi-Ju and Li, Wei and Xi, Bin and Zhang, Zhe and Su, Gang},
  journal = {Phys. Rev. B},
  volume = {86},
  issue = {13},
  pages = {134429},
  numpages = {6},
  year = {2012},
  month = {Oct},
  publisher = {American Physical Society},
  doi = {10.1103/PhysRevB.86.134429},
  url = {https://link.aps.org/doi/10.1103/PhysRevB.86.134429}
}

@article{ChertkovChernyak2006,
  title = {Loop calculus in statistical physics and information science},
  author = {Chertkov, Michael and Chernyak, Vladimir Y.},
  journal = {Phys. Rev. E},
  volume = {73},
  issue = {6},
  pages = {065102},
  numpages = {4},
  year = {2006},
  month = {Jun},
  publisher = {American Physical Society},
  doi = {10.1103/PhysRevE.73.065102},
  url = {https://link.aps.org/doi/10.1103/PhysRevE.73.065102}
}

@article{ChertkovChernyak2006b,
doi = {10.1088/1742-5468/2006/06/P06009},
url = {https://doi.org/10.1088/1742-5468/2006/06/P06009},
year = {2006},
month = {jun},
publisher = {},
volume = {2006},
number = {06},
pages = {P06009},
author = {Chertkov, Michael and Chernyak, Vladimir Y},
title = {Loop series for discrete statistical models on graphs},
journal = {Journal of Statistical Mechanics: Theory and Experiment},
}

@article{Lubasch2014prb,
  title = {Algorithms for finite projected entangled pair states},
  author = {Lubasch, Michael and Cirac, J. Ignacio and Ba\~nuls, Mari-Carmen},
  journal = {Phys. Rev. B},
  volume = {90},
  issue = {6},
  pages = {064425},
  numpages = {16},
  year = {2014},
  month = {Aug},
  publisher = {American Physical Society},
  doi = {10.1103/PhysRevB.90.064425},
  url = {https://link.aps.org/doi/10.1103/PhysRevB.90.064425}
}

@article{gray2018quimb,
  doi = {10.21105/joss.00819},
  url = {https://doi.org/10.21105/joss.00819},
  year = {2018},
  publisher = {The Open Journal},
  volume = {3},
  number = {29},
  pages = {819},
  author = {Johnnie Gray},
  title = {quimb: A python package for quantum information and many-body calculations},
  journal = {Journal of Open Source Software}
}

@misc{symmray,
  author = {Johnnie Gray},
  title = {\texttt{symmray} - a minimal library for block sparse, abelian symetric and fermionic arrays},
  year = {2025},
  publisher = {GitHub},
  journal = {GitHub repository},
  howpublished = {\url{https://github.com/jcmgray/symmray}},
}

@article{gray2021hyper,
  doi = {10.22331/q-2021-03-15-410},
  url = {https://doi.org/10.22331/q-2021-03-15-410},
  title = {Hyper-optimized tensor network contraction},
  author = {Gray, Johnnie and Kourtis, Stefanos},
  journal = {{Quantum}},
  issn = {2521-327X},
  publisher = {{Verein zur F{\"{o}}rderung des Open Access Publizierens in den Quantenwissenschaften}},
  volume = {5},
  pages = {410},
  month = mar,
  year = {2021}
}

@article{Jordaorus2008,
  title = {Classical Simulation of Infinite-Size Quantum Lattice Systems in Two Spatial Dimensions},
  author = {Jordan, J. and Or\'us, R. and Vidal, G. and Verstraete, F. and Cirac, J. I.},
  journal = {Phys. Rev. Lett.},
  volume = {101},
  issue = {25},
  pages = {250602},
  numpages = {4},
  year = {2008},
  month = {Dec},
  publisher = {American Physical Society},
  doi = {10.1103/PhysRevLett.101.250602},
  url = {https://link.aps.org/doi/10.1103/PhysRevLett.101.250602}
}

@book{ran2020tensor,
  title        = {Tensor Network Contractions: Methods and Applications to Quantum Many-Body Systems},
  author       = {Shi-Ju Ran and Emanuele Tirrito and Cheng Peng and Xi Chen and Luca Tagliacozzo and Gang Su and Maciej Lewenstein},
  year         = {2020},
  publisher    = {Springer Cham},
  series       = {Lecture Notes in Physics},
  volume       = {964},
  isbn         = {978-3-030-34488-7, 978-3-030-34489-4},
  doi          = {10.1007/978-3-030-34489-4},
  pages        = {xiv + 150},
  url = {https://doi.org/10.1007/978-3-030-34489-4}
}

@article{ORUS2014117,
title = {A practical introduction to tensor networks: Matrix product states and projected entangled pair states},
journal = {Annals of Physics},
volume = {349},
pages = {117-158},
year = {2014},
issn = {0003-4916},
doi = {https://doi.org/10.1016/j.aop.2014.06.013},
url = {https://www.sciencedirect.com/science/article/pii/S0003491614001596},
author = {Román Orús}
}

@incollection{pearl2022reverend,
  title={Reverend Bayes on inference engines: A distributed hierarchical approach},
  author={Pearl, Judea},
  booktitle={Probabilistic and causal inference: the works of Judea Pearl},
  pages={129--138},
  year={2022},
  doi = {10.1145/3501714.3501727},
  url = {https://doi.org/10.1145/3501714.3501727}
}

@article{albuquerque2007alps,
  title={The ALPS project release 1.3: Open-source software for strongly correlated systems},
  author={Albuquerque, A Fabricio and Alet, Fabien and Corboz, Philippe and Dayal, Prakash and Feiguin, Adrian and Fuchs, Sebastian and Gamper, L and Gull, Emanuel and G{\"u}rtler, S and Honecker, Andreas and others},
  journal={Journal of Magnetism and Magnetic Materials},
  volume={310},
  number={2},
  pages={1187--1193},
  year={2007},
  publisher={Elsevier},
  doi = {10.1016/j.jmmm.2006.10.304},
  url = {https://doi.org/10.1016/j.jmmm.2006.10.304}
}

@article{bauer2011alps,
  title={The ALPS project release 2.0: open source software for strongly correlated systems},
  author={Bauer, Bela and Carr, LD and Evertz, Hans Gerd and Feiguin, Adrian and Freire, J and Fuchs, Sebastian and Gamper, Lukas and Gukelberger, Jan and Gull, Emanuel and Guertler, Siegfried and others},
  journal={Journal of Statistical Mechanics: Theory and Experiment},
  volume={2011},
  number={05},
  pages={P05001},
  year={2011},
  publisher={IOP Publishing},
  doi = {10.1088/1742-5468/2011/05/P05001},
  url = {https://doi.org/10.1088/1742-5468/2011/05/P05001}
}

@article{motta2018ab,
  title={Ab initio computations of molecular systems by the auxiliary-field quantum Monte Carlo method},
  author={Motta, Mario and Zhang, Shiwei},
  journal={Wiley Interdisciplinary Reviews: Computational Molecular Science},
  volume={8},
  number={5},
  pages={e1364},
  year={2018},
  publisher={Wiley Online Library},
  doi = {10.1002/wcms.1364},
  url = {https://doi.org/10.1002/wcms.1364}
}

@article{qin2016benchmark,
  title={Benchmark study of the two-dimensional Hubbard model with auxiliary-field quantum Monte Carlo method},
  author={Qin, Mingpu and Shi, Hao and Zhang, Shiwei},
  journal={Physical Review B},
  volume={94},
  number={8},
  pages={085103},
  year={2016},
  publisher={APS},
  doi = {10.1103/PhysRevB.94.085103},
  url = {https://link.aps.org/doi/10.1103/PhysRevB.94.085103}
}

@article{todo2001cluster,
  title={Cluster algorithms for general-S quantum spin systems},
  author={Todo, Synge and Kato, Kiyoshi},
  journal={Physical review letters},
  volume={87},
  number={4},
  pages={047203},
  year={2001},
  publisher={APS},
  doi = {10.1103/PhysRevLett.87.047203},
  url = {https://link.aps.org/doi/10.1103/PhysRevLett.87.047203}
}

@misc{ad_afqmc,
  author = {Mahajan, Ankit},
  title  = {{AD-AFQMC}: Automatically differentiable {AFQMC}},
  url    = {https://github.com/ankit76/ad_afqmc}
}

@article{mahajan2023response,
  title={Response properties in phaseless auxiliary field quantum Monte Carlo},
  author={Mahajan, Ankit and Kurian, Jo S and Lee, Joonho and Reichman, David R and Sharma, Sandeep},
  journal={The Journal of Chemical Physics},
  volume={159},
  number={18},
  year={2023},
  publisher={AIP Publishing},
  doi = {10.1063/5.0171996},
  url = {https://doi.org/10.1063/5.0171996}
}

@misc{tn_loop_cluster_code_zenodo,
  author    = {Gray, Johnnie},
  title     = {Code and data for ``{Tensor} Network Loop Cluster Expansions
               for Quantum Many-Body Problems''},
  year      = {2026},
  doi       = {10.5281/zenodo.19601138},
  url       = {https://doi.org/10.5281/zenodo.19601138},
  publisher = {Zenodo},
}

\end{document}